\documentclass[a4paper,zpreprint,zbstepj,british]{./zeus_Kpaper}
\usepackage{./zeusdet_hepunits}

\chardef\usc=95
\chardef\til=126
\catcode`\@=11 
\DeclareRobustCommand\xdotspace{\futurelet\@let@token\@xdotspace}
\def\@xdotspace{%
  \ifx\@let@token.\else
  \ifx\@let@token\bgroup.\else
  \ifx\@let@token\egroup.\else
  \ifx\@let@token\/.\else
  \ifx\@let@token\ .\else
  \ifx\@let@token~.\else
  \ifx\@let@token!.\else
  \ifx\@let@token,.\else
  \ifx\@let@token:.\else
  \ifx\@let@token;.\else
  \ifx\@let@token?.\else
  \ifx\@let@token/.\else
  \ifx\@let@token'.\else
  \ifx\@let@token).\else
  \ifx\@let@token-.\else
  \ifx\@let@token\@xobeysp.\else
  \ifx\@let@token\space.\else
  \ifx\@let@token\@sptoken.\else
   .\space
   \fi\fi\fi\fi\fi\fi\fi\fi\fi\fi\fi\fi\fi\fi\fi\fi\fi\fi}
\catcode`\@=12 

\newcommand{\stru}[2]{%
   \relax\ifmmode\hbox{\vrule height#1 depth#2 width0pt}%
   \else\vrule height#1 depth#2 width0pt\fi}

\newcommand{\Ronum}[1]{\uppercase\expandafter{\romannumeral#1}}
\newcommand{\ronum}[1]{\expandafter{\romannumeral#1}}
\DeclareRobustCommand{\LaTeXZ}{%
  \LaTeX\kern-.05em4\kern-.1em
  {\raisebox{-0.2ex}{$\scriptstyle\text{ZEUS}$}}\xspace}

\newcommand{\slashfrac}[2]{%
  \raisebox{0.5ex}{\ensuremath #1}\kern-0.12em/\kern-0.08em
  \raisebox{-.8ex}{\ensuremath #2}}

\newcommand{\sqr}[3]{%
    {\vcenter{\hrule height.#3ex\hbox{\vrule width.#2ex height#1ex
     \kern#1ex\vrule width.#3ex}\hrule height.#2ex}}}

\catcode`\@=11 
\newcommand{\parenbar}{\mathpalette\p@renb@r}
\def\p@renb@r#1#2{\vbox{%
  \ifx#1\scriptscriptstyle \dimen@.7em\dimen@ii.2em\else
  \ifx#1\scriptstyle \dimen@.8em\dimen@ii.25em\else
  \dimen@1em\dimen@ii.4em\fi\fi \offinterlineskip
  \ialign{\hfill##\hfill\cr
    \vbox{\hrule width\dimen@ii}\cr
    \noalign{\vskip-.3ex}%
    \hbox to\dimen@{$\mathchar300\hfil\mathchar301$}\cr
    \noalign{\vskip-.3ex}%
    $#1#2$\cr}}}
\catcode`\@=12 

\ifzeusunit
  
\else
  
\fi

\newcommand{\IP}{{\rm I$\kern-0.01667em$P}\xspace}

\mathchardef\qsm=63
\mathchardef\pls=43
\mathchardef\mns=512
\mathchardef\plm=518
\mathchardef\eql=61
\mathchardef\smallleft=300
\mathchardef\smallright=301
\mathchardef\les=316
\mathchardef\gre=318
\mathchardef\leq=532
\mathchardef\grq=533

\catcode`\@=11 
\newcounter{pict@width}
\newcounter{pict@height}
\newlength{\pict@scale}
\newcommand{\psfigadd}[4]{%
\setcounter{pict@width}{1*\ratio{#2+\pict@scale/2}{\pict@scale}}
\setcounter{pict@height}{1*\ratio{#3+\pict@scale/2}{\pict@scale}}
\setlength{\unitlength}{\pict@scale}
\hbox to #2{\hspace{-\fill}\begin{picture}(\thepict@width,\thepict@height)
\put(0,0){\psfig{figure=#1,width=#2,height=#3,clip=}}
\SetScale{0.283466457}
\SetWidth{1.763889}
{#4}
\end{picture}}
}
\newcounter{pict@widthfst}
\newcounter{pict@widthscd}
\newcounter{pict@widthtot}
\newcommand{\psfigaddtwo}[7]{%
\setcounter{pict@widthfst}{1*\ratio{#2+\pict@scale/2}{\pict@scale}}
\setcounter{pict@widthscd}{1*\ratio{#2+#4+\pict@scale/2}{\pict@scale}}
\setcounter{pict@widthtot}{1*\ratio{#2+#4+#6+\pict@scale/2}{\pict@scale}}
\setcounter{pict@height}{1*\ratio{#3+\pict@scale/2}{\pict@scale}}
\setlength{\unitlength}{\pict@scale}
\hbox{\hspace{-\fill}\begin{picture}(\thepict@widthtot,\thepict@height)
\put(0,0){\psfig{figure=#1,width=#2,height=#3,clip=}}
\put(\thepict@widthscd,0){\psfig{figure=#5,width=#6,height=#3,clip=}}
\SetScale{0.283466457}
\SetWidth{1.763889}
{#7}
\end{picture}}
}
\newcommand{\psfigror}[4]{%
\setcounter{pict@width}{1*\ratio{#2+\pict@scale/2}{\pict@scale}}
\setcounter{pict@height}{1*\ratio{#3+\pict@scale/2}{\pict@scale}}
\setlength{\unitlength}{\pict@scale}
\hbox{\begin{picture}(\thepict@width,\thepict@height)
\put(0,\thepict@height){\psfig{figure=#1,width=#3,height=#2,clip=,angle=270}}
\SetScale{0.283466457}
\SetWidth{1.763889}
{#4}
\end{picture}}
}
\newcommand{\psfigrol}[4]{%
\setcounter{pict@width}{1*\ratio{#2+\pict@scale/2}{\pict@scale}}
\setcounter{pict@height}{1*\ratio{#3+\pict@scale/2}{\pict@scale}}
\setlength{\unitlength}{\pict@scale}
\hbox{\begin{picture}(\thepict@width,\thepict@height)
\put(0,0){\psfig{figure=#1,width=#3,height=#2,clip=,angle=90}}
\SetScale{0.283466457}
\SetWidth{1.763889}
{#4}
\end{picture}}
}
\catcode`\@=12 
\newlength\listtextwidth

\catcode`\@=11 
\newlength{\@tabfninsert}
\newlength{\@tabfnwidth}
\newcommand{\tabfootnote}[2]{%
  \setlength{\@tabfninsert}{0.8em}
  \setlength{\@tabfnwidth}{\textwidth}
  \addtolength{\@tabfnwidth}{-\@tabfninsert}
  \addtolength{\@tabfnwidth}{-0.4em}
  \noindent\makebox[\@tabfninsert][r]{\footnotesize$^{#1}$\hfil}\hfill%
  \parbox[t]{\@tabfnwidth}{\footnotesize #2\hfill}}
\catcode`\@=12 

\newcommand{\qq}{q\overline{q}}
\newcommand{\qqg}{q\overline{q}g}
\newcommand{\xPom}{x_{\textrm{I\!P}}}
\newcommand{\jetRM}{\textrm{jet}}
\newcommand{\fpdiss}{f_{\textrm{pdiss}}}
\newcommand{\etaMax}{\eta_{\textrm{max}}}

\newcommand{\ptcut}{p_{T,\textrm{cut}}}
\newcommand{\phiCross}{\textrm{d}\sigma/\textrm{d}\phi}
\newcommand{\betaCross}{\textrm{d}\sigma/\textrm{d}\beta}
\newcommand{\ptjet}{p_{T,\textrm{jet}}}
\newcommand{\cosTwoPhi}{\cos 2\phi}
\newcommand{\myFig}{Fig.}
\newcommand{\myFigs}{Figs.}
\newcommand{\myFigure}{Figure}

\def\citeCTD{{\cite{%
nim:a279:290,*npps:b32:181,*nim:a338:254%
}}\xspace}
\def\citeMVD{{\cite{%
nim:a581:656%
}}\xspace}
\def\citeSTT{{\cite{%
nim:a535:191%
}}\xspace}
\def\citeCAL{{\cite{%
nim:a309:77,*nim:a309:101,*nim:a321:356,*nim:a336:23%
}}\xspace}
\def\citeHES{{\cite{%
nim:a277:176%
}}\xspace}
\def\citeSixmT{{\cite{%
thesis:gosau:2007%
}}\xspace}
\def\citeBAC{{\cite{%
nim:a300:480%
}}\xspace}
\def\citePCAL{{\cite{%
desy-92-066,*zfp:c63:391,*acpp:b32:2025%
}}\xspace}
\def\citeSPECTRO{{\cite{%
nim:a565:572%
}}\xspace}

\begin{document}
\graphicspath{{./}}

%
%
\prepnum{DESY-15-070}
\prepdate{May 2015}

\zeustitle{%
Production of exclusive dijets in diffractive deep inelastic scattering at HERA
}
                    
\zeusauthor{ZEUS Collaboration}

\draftversion{4.1}
\zeusdate{}

\maketitle

%
%
\begin{abstract}\noindent
  Production of exclusive dijets in diffractive deep inelastic $e^\pm
  p$ scattering has been measured with the ZEUS detector at HERA using
  an integrated luminosity of \linebreak \unit{372}{\invpb}.  The
  measurement was performed for $\gamma^*$--$p$ centre-of-mass
  energies in the range $90 < W < \unit{250}{\GeV}$ and for photon
  virtualities $Q^2 > \unit{25}{\GeV\squared}$.  Energy flows around
  the jet axis are presented. The cross section is presented as a
  function of $\beta$ and $\phi$, where $\beta=x/x_{\rm I\!P}$, $x$ is
  the Bjorken variable and $x_{\rm I\!P}$ is the proton fractional
  longitudinal momentum loss. The angle $\phi$ is defined by the
  \mbox{$\gamma^*$--dijet} plane and the $\gamma^*$--$e^\pm$ plane in
  the rest frame of the diffractive final state. The $\phi$ cross
  section is measured in bins of $\beta$. The results are compared to
  predictions from models based on different assumptions about the
  nature of the diffractive exchange.

\end{abstract}

\thispagestyle{empty}

%
%
%
%

\topmargin-1.cm
\evensidemargin-0.3cm
\oddsidemargin-0.3cm
\textwidth 16.cm
\textheight 680pt
\parindent0.cm
\parskip0.3cm plus0.05cm minus0.05cm
\def\3{\ss}
\pagenumbering{Roman}
                                                   %
\begin{center}
{                      \Large  The ZEUS Collaboration              }
\end{center}

{\small\raggedright


H.~Abramowicz$^{26, u}$, 
I.~Abt$^{21}$, 
L.~Adamczyk$^{8}$, 
M.~Adamus$^{32}$, 
S.~Antonelli$^{2}$, 
V.~Aushev$^{16, 17, o}$, 
Y.~Aushev$^{17, o, p}$, 
O.~Behnke$^{10}$, 
U.~Behrens$^{10}$, 
A.~Bertolin$^{23}$
% ...
\newpage


{\setlength{\parskip}{0.4em}
\makebox[3ex]{$^{1}$}
\begin{minipage}[t]{14cm}
{\it INFN Bologna, Bologna, Italy}~$^{A}$

\end{minipage}

\makebox[3ex]{$^{2}$}
\begin{minipage}[t]{14cm}
{\it University and INFN Bologna, Bologna, Italy}~$^{A}$

\end{minipage}

\makebox[3ex]{$^{3}$}
\begin{minipage}[t]{14cm}
{\it Physikalisches Institut der Universit\"at Bonn,
Bonn, Germany}~$^{B}$

\end{minipage}

\makebox[3ex]{$^{4}$}
\begin{minipage}[t]{14cm}
{\it Panjab University, Department of Physics, Chandigarh, India}

\end{minipage}

\makebox[3ex]{$^{5}$}
\begin{minipage}[t]{14cm}
{\it Calabria University,
Physics Department and INFN, Cosenza, Italy}~$^{A}$

\end{minipage}

\makebox[3ex]{$^{6}$}
\begin{minipage}[t]{14cm}
{\it National Centre for Particle Physics, Universiti Malaya, 50603 Kuala Lumpur, Malaysia}~$^{C}$

\end{minipage}

\makebox[3ex]{$^{7}$}
\begin{minipage}[t]{14cm}
{\it The Henryk Niewodniczanski Institute of Nuclear Physics, Polish Academy of \\
Sciences, Krakow, Poland}~$^{D}$

\end{minipage}

\makebox[3ex]{$^{8}$}
\begin{minipage}[t]{14cm}
{\it AGH-University of Science and Technology, Faculty of Physics and Applied Computer
Science, Krakow, Poland}~$^{D}$

\end{minipage}

\makebox[3ex]{$^{9}$}
\begin{minipage}[t]{14cm}
{\it Department of Physics, Jagellonian University, Krakow, Poland}

\end{minipage}

\makebox[3ex]{$^{10}$}
\begin{minipage}[t]{14cm}
{\it Deutsches Elektronen-Synchrotron DESY, Hamburg, Germany}

\end{minipage}

\makebox[3ex]{$^{11}$}
\begin{minipage}[t]{14cm}
{\it Deutsches Elektronen-Synchrotron DESY, Zeuthen, Germany}

\end{minipage}

\makebox[3ex]{$^{12}$}
\begin{minipage}[t]{14cm}
{\it School of Physics and Astronomy, University of Glasgow,
Glasgow, United Kingdom}~$^{E}$

\end{minipage}

\makebox[3ex]{$^{13}$}
\begin{minipage}[t]{14cm}
{\it Hamburg University, Institute of Experimental Physics, Hamburg,
Germany}~$^{F}$

\end{minipage}

\makebox[3ex]{$^{14}$}
\begin{minipage}[t]{14cm}
{\it Institute of Particle and Nuclear Studies, KEK,
Tsukuba, Japan}~$^{G}$

\end{minipage}

\makebox[3ex]{$^{15}$}
\begin{minipage}[t]{14cm}
{\it Institute of Physics and Technology of Ministry of Education and
Science of Kazakhstan, Almaty, Kazakhstan}

\end{minipage}

\makebox[3ex]{$^{16}$}
\begin{minipage}[t]{14cm}
{\it Institute for Nuclear Research, National Academy of Sciences, Kyiv, Ukraine}

\end{minipage}

\makebox[3ex]{$^{17}$}
\begin{minipage}[t]{14cm}
{\it Department of Nuclear Physics, National Taras Shevchenko University of Kyiv, Kyiv, Ukraine}

\end{minipage}

\makebox[3ex]{$^{18}$}
\begin{minipage}[t]{14cm}
{\it Department of Physics, McGill University,
Montr\'eal, Qu\'ebec, Canada H3A 2T8}~$^{H}$

\end{minipage}

\makebox[3ex]{$^{19}$}
\begin{minipage}[t]{14cm}
{\it Meiji Gakuin University, Faculty of General Education,
Yokohama, Japan}~$^{G}$

\end{minipage}

\makebox[3ex]{$^{20}$}
\begin{minipage}[t]{14cm}
{\it Lomonosov Moscow State University, Skobeltsyn Institute of Nuclear Physics,
Moscow, Russia}~$^{I}$

\end{minipage}

\makebox[3ex]{$^{21}$}
\begin{minipage}[t]{14cm}
{\it Max-Planck-Institut f\"ur Physik, M\"unchen, Germany}

\end{minipage}

\makebox[3ex]{$^{22}$}
\begin{minipage}[t]{14cm}
{\it Department of Physics, University of Oxford,
Oxford, United Kingdom}~$^{E}$

\end{minipage}

\makebox[3ex]{$^{23}$}
\begin{minipage}[t]{14cm}
{\it INFN Padova, Padova, Italy}~$^{A}$

\end{minipage}

\makebox[3ex]{$^{24}$}
\begin{minipage}[t]{14cm}
{\it Dipartimento di Fisica e Astronomia dell' Universit\`a and INFN,
Padova, Italy}~$^{A}$

\end{minipage}

\makebox[3ex]{$^{25}$}
\begin{minipage}[t]{14cm}
{\it Polytechnic University, Tokyo, Japan}~$^{G}$

\end{minipage}

\makebox[3ex]{$^{26}$}
\begin{minipage}[t]{14cm}
{\it Raymond and Beverly Sackler Faculty of Exact Sciences, School of Physics, \\
Tel Aviv University, Tel Aviv, Israel}~$^{J}$

\end{minipage}

\makebox[3ex]{$^{27}$}
\begin{minipage}[t]{14cm}
{\it Department of Physics, Tokyo Institute of Technology,
Tokyo, Japan}~$^{G}$

\end{minipage}

\makebox[3ex]{$^{28}$}
\begin{minipage}[t]{14cm}
{\it Universit\`a di Torino and INFN, Torino, Italy}~$^{A}$

\end{minipage}

\makebox[3ex]{$^{29}$}
\begin{minipage}[t]{14cm}
{\it Universit\`a del Piemonte Orientale, Novara, and INFN, Torino,
Italy}~$^{A}$

\end{minipage}

\makebox[3ex]{$^{30}$}
\begin{minipage}[t]{14cm}
{\it Physics and Astronomy Department, University College London,
London, United Kingdom}~$^{E}$

\end{minipage}

\makebox[3ex]{$^{31}$}
\begin{minipage}[t]{14cm}
{\it Faculty of Physics, University of Warsaw, Warsaw, Poland}

\end{minipage}

\makebox[3ex]{$^{32}$}
\begin{minipage}[t]{14cm}
{\it National Centre for Nuclear Research, Warsaw, Poland}

\end{minipage}

\makebox[3ex]{$^{33}$}
\begin{minipage}[t]{14cm}
{\it Department of Particle Physics and Astrophysics, Weizmann
Institute, Rehovot, Israel}

\end{minipage}

\makebox[3ex]{$^{34}$}
\begin{minipage}[t]{14cm}
{\it Department of Physics, York University, Ontario, Canada M3J 1P3}~$^{H}$

\end{minipage}

}

\vspace{3em}


{\setlength{\parskip}{0.4em}\raggedright
\makebox[3ex]{$^{ A}$}
\begin{minipage}[t]{14cm}
 supported by the Italian National Institute for Nuclear Physics (INFN) \
\end{minipage}

\makebox[3ex]{$^{ B}$}
\begin{minipage}[t]{14cm}
 supported by the German Federal Ministry for Education and Research (BMBF), under
 contract No. 05 H09PDF\
\end{minipage}

\makebox[3ex]{$^{ C}$}
\begin{minipage}[t]{14cm}
 supported by HIR grant UM.C/625/1/HIR/149 and UMRG grants RU006-2013, RP012A-13AFR and RP012B-13AFR from
 Universiti Malaya, and ERGS grant ER004-2012A from the Ministry of Education, Malaysia\
\end{minipage}

\makebox[3ex]{$^{ D}$}
\begin{minipage}[t]{14cm}
 supported by the National Science Centre under contract No. DEC-2012/06/M/ST2/00428\
\end{minipage}

\makebox[3ex]{$^{ E}$}
\begin{minipage}[t]{14cm}
 supported by the Science and Technology Facilities Council, UK\
\end{minipage}

\makebox[3ex]{$^{ F}$}
\begin{minipage}[t]{14cm}
 supported by the German Federal Ministry for Education and Research (BMBF), under
 contract No. 05h09GUF, and the SFB 676 of the Deutsche Forschungsgemeinschaft (DFG) \
\end{minipage}

\makebox[3ex]{$^{ G}$}
\begin{minipage}[t]{14cm}
 supported by the Japanese Ministry of Education, Culture, Sports, Science and Technology
 (MEXT) and its grants for Scientific Research\
\end{minipage}

\makebox[3ex]{$^{ H}$}
\begin{minipage}[t]{14cm}
 supported by the Natural Sciences and Engineering Research Council of Canada (NSERC) \
\end{minipage}

\makebox[3ex]{$^{ I}$}
\begin{minipage}[t]{14cm}
 supported by RF Presidential grant N 3042.2014.2 for the Leading Scientific Schools and by
 the Russian Ministry of Education and Science through its grant for Scientific Research on
 High Energy Physics\
\end{minipage}

\makebox[3ex]{$^{ J}$}
\begin{minipage}[t]{14cm}
 supported by the Israel Science Foundation\
\end{minipage}

}

\pagebreak[4]
{\setlength{\parskip}{0.4em}


\makebox[3ex]{$^{ a}$}
\begin{minipage}[t]{14cm}
also funded by Max Planck Institute for Physics, Munich, Germany, now at Sant Longowal Institute of Engineering and Technology, Longowal, Punjab, India\
\end{minipage}

\makebox[3ex]{$^{ b}$}
\begin{minipage}[t]{14cm}
also funded by Max Planck Institute for Physics, Munich, Germany, now at Sri Guru Granth Sahib World University, Fatehgarh Sahib, India\
\end{minipage}

\makebox[3ex]{$^{ c}$}
\begin{minipage}[t]{14cm}
also at Agensi Nuklear Malaysia, 43000 Kajang, Bangi, Malaysia\
\end{minipage}

\makebox[3ex]{$^{ d}$}
\begin{minipage}[t]{14cm}
now at School of Physics and Astronomy, University of Birmingham, UK\
\end{minipage}

\makebox[3ex]{$^{ e}$}
\begin{minipage}[t]{14cm}
partially supported by the Polish National Science Centre projects DEC-2011/01/B/ST2/03643 and DEC-2011/03/B/ST2/00220\
\end{minipage}

\makebox[3ex]{$^{ f}$}
\begin{minipage}[t]{14cm}
now at Rockefeller University, New York, NY 10065, USA\
\end{minipage}

\makebox[3ex]{$^{ g}$}
\begin{minipage}[t]{14cm}
now at European X-ray Free-Electron Laser facility GmbH, Hamburg, Germany\
\end{minipage}

\makebox[3ex]{$^{ h}$}
\begin{minipage}[t]{14cm}
now at University of Liverpool, United Kingdom\
\end{minipage}

\makebox[3ex]{$^{ i}$}
\begin{minipage}[t]{14cm}
supported by the Alexander von Humboldt Foundation\
\end{minipage}

\makebox[3ex]{$^{ j}$}
\begin{minipage}[t]{14cm}
now at CERN, Geneva, Switzerland\
\end{minipage}

\makebox[3ex]{$^{ k}$}
\begin{minipage}[t]{14cm}
Alexander von Humboldt Professor; also at DESY and University of Oxford\
\end{minipage}

\makebox[3ex]{$^{ l}$}
\begin{minipage}[t]{14cm}
also at DESY\
\end{minipage}

\makebox[3ex]{$^{ m}$}
\begin{minipage}[t]{14cm}
also at University of Tokyo, Japan\
\end{minipage}

\makebox[3ex]{$^{ n}$}
\begin{minipage}[t]{14cm}
now at Kobe University, Japan\
\end{minipage}

\makebox[3ex]{$^{ o}$}
\begin{minipage}[t]{14cm}
supported by DESY, Germany\
\end{minipage}

\makebox[3ex]{$^{ p}$}
\begin{minipage}[t]{14cm}
member of National Technical University of Ukraine, Kyiv Polytechnic Institute, Kyiv, Ukraine\
\end{minipage}

\makebox[3ex]{$^{ q}$}
\begin{minipage}[t]{14cm}
now at DESY ATLAS group\
\end{minipage}

\makebox[3ex]{$^{ r}$}
\begin{minipage}[t]{14cm}
member of National University of Kyiv - Mohyla Academy, Kyiv, Ukraine\
\end{minipage}

\makebox[3ex]{$^{ s}$}
\begin{minipage}[t]{14cm}
now at BNL, USA\
\end{minipage}

\makebox[3ex]{$^{ t}$}
\begin{minipage}[t]{14cm}
now at LNF, Frascati, Italy\
\end{minipage}

\makebox[3ex]{$^{ u}$}
\begin{minipage}[t]{14cm}
also at Max Planck Institute for Physics, Munich, Germany, External Scientific Member\
\end{minipage}

\makebox[3ex]{$^{ v}$}
\begin{minipage}[t]{14cm}
also at Universit\"{a}t Hamburg and supported by DESY and the Alexander von Humboldt Foundation\
\end{minipage}

\makebox[3ex]{$^{ w}$}
\begin{minipage}[t]{14cm}
also at \L\'{o}d\'{z} University, Poland\
\end{minipage}

\makebox[3ex]{$^{ x}$}
\begin{minipage}[t]{14cm}
member of \L\'{o}d\'{z} University, Poland\
\end{minipage}

\makebox[3ex]{$^{ y}$}
\begin{minipage}[t]{14cm}
now at Polish Air Force Academy in Deblin\
\end{minipage}

}

}

\cleardoublepage
\cleardoublepage

\pagenumbering{arabic}
\section{Introduction}
\label{sec-int}

The first evidence for exclusive dijet production at high-energy
hadron colliders was provided by the CDF experiment at the Fermilab
Tevatron $p\bar{p}$ collider~\cite{Aaltonen:2007hs} and had an
important impact on theoretical calculations of exclusive Higgs boson
production at the Large Hadron Collider. This paper describes the
first measurement of exclusive dijet production in high energy
electron\footnote{Here and in the following the term ``electron''
  denotes generically both the electron and the positron.}--proton
scattering. A quantitative understanding of the production of
exclusive dijets in lepton--hadron scattering can improve the 
understanding of more complicated processes like the exclusive
production of dijets in hadron--hadron scattering~\cite{Martin:1997kv}
or in lepton--ion scattering at a future eRHIC
accelerator~\cite{Goloskokov:2004br}.

\begin{figure}
\centering
\includegraphics[height=0.3\textheight]{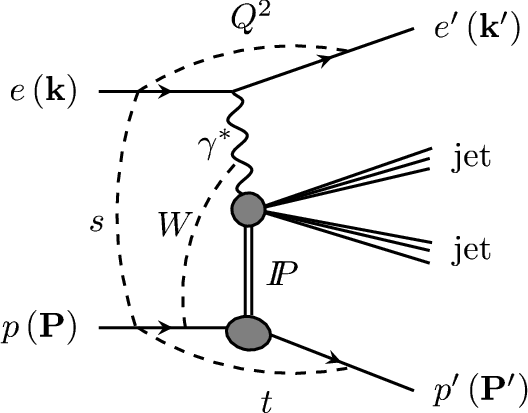}
\caption[]{Schematic view of the diffractive production of exclusive dijets in electron--proton DIS.}   
\label{fig_kinematics}
\end{figure}

\begin{figure}
\centering
\includegraphics[height=0.46\textheight]{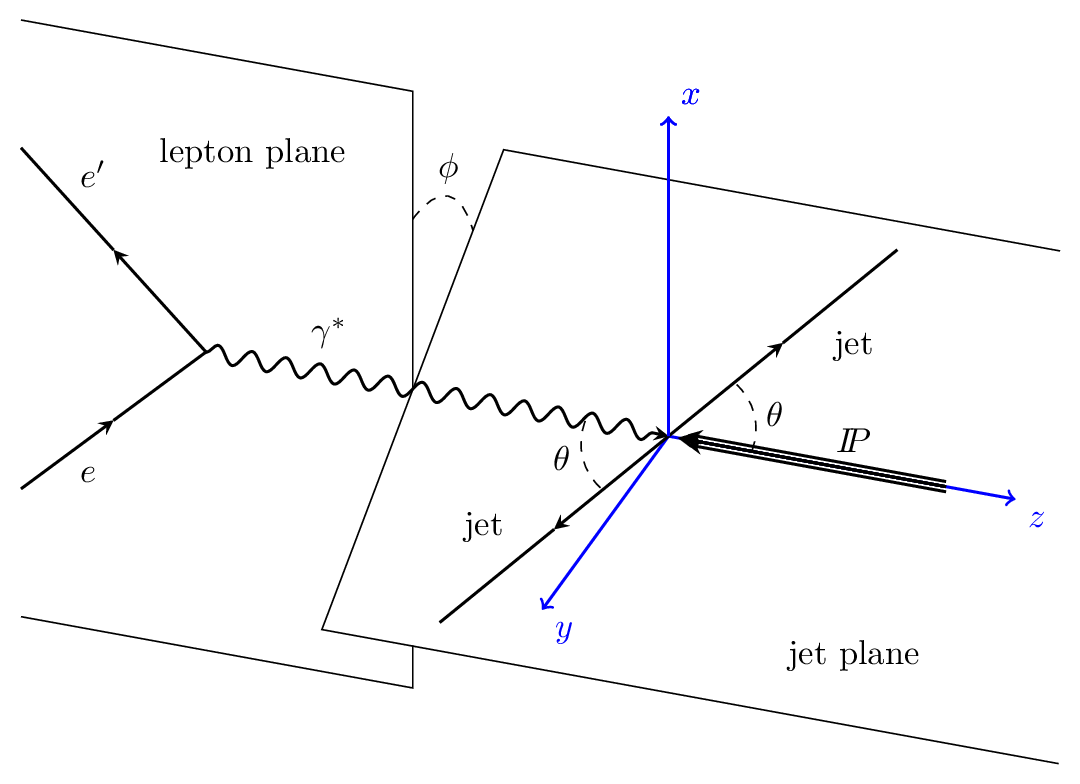}
\caption[]{Definition of planes and angles in the $\gamma^*$--${\rm
    I\!P}$ centre-of-mass system. The lepton plane is defined by the
  $\gamma^*$ and $e$ momenta.  The jet plane is defined by the
  $\gamma^*$ and dijet directions. The angle $\phi$ is the angle
  between these two planes.  The jet polar angle, $\theta$, is the
  angle between the directions of the jets and $\gamma^*$.  }
\label{fig_azim_angle_def}
\end{figure}

A schematic view of the diffractive production of exclusive dijets, $e
+ p \rightarrow e+ \mathrm{jet}1 + \mathrm{jet}2 + p$, is shown in
\myFig{}~\ref{fig_kinematics}. In this picture, electron--proton deep
inelastic scattering (DIS) is described in terms of an interaction
between the virtual photon, $\gamma^*$, and the proton, which is
mediated by the exchange of a colourless object called the Pomeron
(${\rm I\!P}$). This process in the $\gamma^*$--${\rm I\!P}$
centre-of-mass frame is presented in
\myFig{}~\ref{fig_azim_angle_def}, where the lepton and jet planes
are marked. The lepton plane is defined by the incoming and scattered
electron momenta. The jet plane is defined by the jet momenta, which are
always back-to-back, and the virtual photon momentum. The angle
between these planes is labelled $\phi$. The jet polar angle is defined
with respect to the virtual photon momentum and called $\theta$.

The production of exclusive dijets in DIS is sensitive to the nature
of the object exchanged between the virtual photon and the proton.
Calculations of the single-differential cross section of dijet
production as a function of $\phi$ in
$k_t$-factorisation~\cite{pl:b386:389} and collinear
factorisation~\cite{Braun:2005rg} have shown that, when the quark and
antiquark jets are indistinguishable, the cross section is
proportional to $1 + A(\ptjet) \cosTwoPhi$, where $\ptjet$ is the jet
transverse momentum. It was pointed out for the first time by Bartels
et al.~\cite{pl:b379:239, pl:b386:389} that the parameter $A$ is
positive if the quark--antiquark pair is produced via the interaction
of a single gluon with the virtual photon and negative if a system of
two gluons takes part in the interaction. The absolute value of the
$A$ parameter is expected to increase as the transverse momentum of
the jet increases.

The production of exclusive dijets is also sensitive to the gluon
distribution in the proton and is a promising reaction to probe the
off-diagonal (generalised~\cite{GolecBiernat:1998vf}) gluon
distribution.  The off-diagonal calculations predict a larger cross
section compared to calculations based on conventional gluon
distributions.  In this context, the exclusive production of dijets is
a complementary process to the exclusive production of vector mesons
which has been extensively studied at HERA\cite{Aaron:2009xp,
  Chekanov:2007zr, pl:b487:273, zfp:c73:73, np:b695:3, pl:b377:259,
  Aktas:2005xu, epj:c24:345, Chekanov:2009zz}.

This paper describes the measurement of differential cross sections as
a function of $\beta$ and in bins of $\beta$ as a function of
$\phi$. The former quantity is defined as $\beta=x/x_{\rm I\!P}$,
where $x$ is the Bjorken variable and $x_{\rm I\!P}$ is the fractional
loss of proton longitudinal momentum. The results of this analysis are
compared to predictions from the Two-Gluon-Exchange
model~\cite{pl:b379:239, epj:c11:111} and the Resolved-Pomeron model
of Ingelman and Schlein~\cite {pl:b152:256}.

\section{Experimental set-up}
\label{sec-exp}
\Zdetdesc

In the kinematic range of the analysis, charged particles were tracked
in the central tracking detector (CTD)~\citeCTD and the microvertex
detector (MVD)~\cite{polini:2007}. These components operated in a magnetic
field of \unit{1.43}{\tesla} provided by a thin superconducting solenoid. The
CTD consisted of 72~cylindrical drift-chamber layers, organised in nine
superlayers covering the polar-angle\ZcoosysfnBEeta{} region
\mbox{$\unit{15}{\degree}<\theta<\unit{164}{\degree}$}.
The MVD silicon tracker consisted of a barrel (BMVD) and a forward
(FMVD) section. The BMVD contained three layers and provided
polar-angle coverage for tracks from $\unit{30}{\degree}$ to
$\unit{150}{\degree}$. The four-layer FMVD extended the polar-angle coverage in
the forward region to $\unit{7}{\degree}$. After alignment, the single-hit
resolution of the MVD was \unit{24}{\micron}. The transverse distance of closest
approach (DCA) of tracks to the nominal vertex in $X$--$Y$ was measured to have
a resolution, averaged over the azimuthal angle, of \unit{$(46 \oplus 122 /
p_{T})$}{\micron}, with $p_{T}$ in \GeV.  For CTD--MVD tracks that pass
through all nine CTD superlayers, the momentum resolution was
$\sigma(p_{T})/p_{T} = 0.0029 p_{T} \oplus 0.0081 \oplus
0.0012/p_{T}$, with $p_{T}$ in \GeV.

\Zcaldesc

%
The position of electrons scattered at small angles to the
electron beam direction was determined with the help of RHES\citeHES,
which consisted of a layer of approximately $10\,000$ $(2.96\times
3.32\,{\rm cm^2})$ silicon-pad detectors inserted in the RCAL at a
depth of 3.3 radiation lengths.

%
%
The luminosity was measured using the Bethe--Heitler reaction
$ep\,\rightarrow\, e\gamma p$ by a luminosity detector which consisted
of independent lead--scintillator calorimeter\citePCAL and magnetic
spectrometer\citeSPECTRO systems. The fractional systematic
uncertainty on the measured luminosity was 2\%~\cite{Adamczyk:2013ewk}.

\section{Monte Carlo simulation}
\label{sec:monte-carlo}

Samples of Monte Carlo (MC) events were generated to determine the
response of the detector to jets of hadrons and the correction factors
necessary to obtain the hadron-level jet cross sections.  The hadron
level is defined in terms of hadrons with lifetime $\geq
\unit{10}{\pico\second}$.  The generated events were passed through
the GEANT 3.21-based~\cite{tech:cern-dd-ee-84-1} ZEUS detector- and
trigger-simulation programs~\cite{zeus:1993:bluebook}.  They were
reconstructed and analysed by the same program chain used for real
data.

In this analysis, the model SATRAP~\cite{pr:d59:014017,pr:d60:114023}
as implemented in the RAPGAP~\cite{cpc:86:147} program was used to
generate diffractive events. SATRAP is a colour-dipole
model~\cite{Nikolaev:1993th} which includes saturation effects.  It
describes DIS as a fluctuation of the virtual photon into a
quark--antiquark dipole which scatters off the proton.  The
CTEQ5D~\cite{epj:c12:375} parameterisation was used to describe the
proton structure. Hadronisation was simulated with the JETSET
7.4~\cite{cpc:82:74,manual:cern-th-7112/93} program which is based on
the Lund string model~\cite{prep:97:31}.  Radiative corrections for
initial- and final-state electromagnetic radiation were taken into
account with the HERACLES 4.6.6~\cite{cpc:69:155,cpc:81:381} program.
The diffractive MC was weighted in order to describe the measured
distributions (see Section~\ref{sec:mc-vs-data}).

The proton-dissociation process was modelled using the
EPSOFT~\cite{thesis:kasprzak:1994,thesis:adamczyk:1999} generator.
The production of dijets is not implemented in EPSOFT.  Therefore
dijets with proton dissociation were simulated with SATRAP, where the
intact proton was replaced with a dissociated proton.  Such a solution
is based on the factorisation hypothesis which assumes that the
interaction at the lepton and at the proton vertex factorises. The factorisation
hypothesis has been verified for diffractive processes in $ep$
collisions at HERA~\cite{epj:c2:247,zfp:c75:607,Chekanov:2004hy,Aktas:2006hy}.

To estimate the non-diffractive DIS background, a sample of events was
generated using HERACLES 4.6.6~\cite{cpc:69:155,cpc:81:381} with
DJANGOH 1.6~\cite{spi:www:MYdjangoh} interfaced to the hadronisation
process. The QCD cascade was simulated using the colour-dipole model
(CDM)~\cite{Gustafson:1986db,np:b306:746,Andersson:1988gp} as
implemented in ARIADNE 4.08~\cite{cpc:71:15,Lonnblad:1994wk}.

To estimate the background of diffractive dijet photoproduction, a
sample of events was generated using the PYTHIA
6.2~\cite{hep-ph-0108264} program with the CTEQ4L~\cite{pr:d55:1280}
parton density function of the proton. The hadronisation process was
simulated with JETSET 7.4.

For the model predictions, events were generated using RAPGAP where
both the Resolved-Pomeron model and the Two-Gluon-Exchange model are
implemented. The hadronisation was simulated with ARIADNE. The
generated events do not include proton dissociation.

In this analysis, the number of diffractive MC events was normalised to
the number of events observed in the data after all selection cuts and
after subtraction of background from photoproduction and
non-diffractive DIS. The numbers of background events were estimated
based on generator cross sections.

\section{Event selection and reconstruction}
\label{sec:event-select-reconst}

This analysis is based on data collected with the ZEUS detector at the
HERA collider during the 2003--2007 data-taking period, when electrons
or positrons of $\unit{27.5}{\GeV}$ were collided with protons of
$\unit{920}{\GeV}$ at a centre-of-mass energy of
$\sqrt{s}=\unit{318}{\GeV}$. The data sample corresponds to an
integrated luminosity of \unit{372}{\invpb}.

A three-level trigger system was used to select events online
\cite{zeus:1993:bluebook,epj:c1:109}.  At the first level, only coarse
calorimeter and tracking information were available. Events
consistent with diffractive DIS were selected using criteria based on
the energy and transverse energy measured in the CAL.  At the second
level, charged-particle tracks were reconstructed online by the ZEUS
global tracking trigger~\cite{Smith:1992im,Allfrey:2007zz}, which combined
information from the CTD and MVD. These online tracks were used to
reconstruct the interaction vertex and to reject non-$ep$
background. At the third level, neutral current DIS events were
accepted on the basis of the identification of a scattered electron
candidate using localised energy depositions in the CAL.

The scattered electron was identified using a neural-network
algorithm~\cite{nim:a365:508}. The reconstruction of the scattered
electron variables was based on the information from the CAL. The
energy of electrons hitting the RCAL was corrected for the presence of
dead material using the rear presampler detector~\cite{nim:a382:419}.
Energy-flow objects (EFOs~\cite{thesis:briskin:1998,
  thesis:tuning:2001}) were used to combine the information from the
CAL and the CTD.

\subsection{DIS selection}

A clean sample of DIS events with a well-reconstructed electron was selected by the following criteria:

\begin{itemize}
\item the electron candidate was reconstructed with calorimeter
  information and was required to have energy reconstructed with
  double-angle method~\cite{proc:hera:1991:23}, $E_e^\prime >
  \unit{10}{\GeV}$ and, if reconstructed in the CTD acceptance region,
  also an associated track;
\item the reconstructed position of the electron candidate in the CAL
  was required to be outside the regions of CAL in which the scattered
  electron might have crossed a substantial amount of inactive
  material or regions with poor acceptance;
\item the vertex position along the beam axis was required to be in
  the range $|Z_{\rm{vtx}}| < \unit{30}{\centi\metre}$;
\item $E_{\mathrm{had}}/E_{\mathrm{tot}} > 0.06$, where
  $E_{\mathrm{had}}$ is the energy deposited in the hadronic part of
  the CAL and $E_{\mathrm{tot}}$ is the total energy in the CAL; this
  cut removes purely electromagnetic events;
\item $45 < (E - P_Z) < \unit{70}{\GeV}$, where $E$ is the total
  energy, $E = \sum_i{E_i}$, $P_{Z} = \sum_i{p_{Z,i}}$ and $p_{Z,i} =
  E_i \cos \theta_{i}$, where the sums run over all EFOs including
  the electron; this cut removes events with large initial-state
  radiation and further reduces the background from photoproduction.
\end{itemize}
Events were accepted if $Q^2 > \unit{25}{\GeV\squared}$ and $90 < W <
\unit{250}{\GeV}$. In this analysis, the photon virtuality, $Q^2$, and
the total energy in the virtual-photon--proton system, $W$, were
reconstructed using the double-angle method which was found to be more
precise than other reconstruction methods in the kinematic region of
this measurement~\cite{Gach:thesis}. The inelasticity, $y$, which was
reconstructed with the electron method was limited to the range $0.1<
y < 0.64$. The limits come from the selection criteria applied to
other variables reconstructed with the double-angle method. The use of
two methods to reconstruct DIS kinematic quantities, increases the
purity if the sample.

\subsection{Diffractive selection}

Diffractive events are characterised by a small momentum exchange at
the proton vertex and by the presence of a large rapidity gap (LRG)
between the proton beam direction and the hadronic final state.
Diffractive DIS events were selected by the following additional
criteria:
\begin{itemize}
\item $x_{\rm I\!P} < 0.01$, where $x_{\rm I\!P}$ is the fraction of
  the proton momentum carried by the diffractive exchange, calculated
  according to the formula $x_{\rm I\!P} = \left(Q^2 + M_X^2 \right) /
  \left( Q^2+W^2 \right)$, in which $M_X$ denotes the invariant mass
  of the hadronic state recoiling against the leading proton and was
  reconstructed from the EFOs excluding the scattered electron
  candidate; this cut reduces the non-diffractive background;
\item $\eta_{\rm max} < 2$, where $\eta_{\rm max}$ is defined as the
  pseudorapidity of the most forward EFO, with an energy greater than
  $E_{\mathrm{EFO}}= \unit{400}{\MeV}$; this cut ensures the presence
  of a LRG in the event;
\item $M_X > \unit{5}{\GeV}$; this cut removes events with resonant
  particle production and ensures that there is enough energy in the
  system to create two jets with high transverse momenta.
\end{itemize}
The origin of exclusive dijet events in diffraction is not unique. The
most natural contribution comes from exclusive production of
quark--antiquark pairs, but other contributions, in particular from
quark--antiquark--gluon, are not excluded. It is predicted~\cite{epj:c11:111} that
the ratio of $\qq$ to $\qqg$ production changes significantly with the
parameter $\beta$ (or $M_X$) in contrast to other kinematic
variables. To get insight into the origin of exclusive diffractive
dijet events, the data were analysed as a function of $\beta$,
calculated according to $\beta = Q^2 / \left(Q^2 + M_X^2 \right)$.

\subsection{Jet selection}
\label{sec:jet-selection}

The $k_T$-cluster algorithm known as the Durham jet
algorithm~\cite{Catani:1991hj, np:b406:187}, as implemented in the
FastJet package~\cite{Cacciari:2011ma}, was used for jet
reconstruction.  Exclusive jets are of interest in this analysis,
so the algorithm was used in ``exclusive mode'' i.e.\ each object
representing a particle or a group of particles had to be finally
associated to a jet. The algorithm is defined in the following way:
first all objects were boosted to the $\gamma^*$--${\rm I\!P}$ rest
frame. Then, the relative distance of each pair of objects,
$k_{T\,ij}^2$, was calculated as \[ k_{T\,ij}^2 = 2\, {\rm
  min}(E_i^2,E_j^2)(1-\cos \theta_{ij}), \] where $\theta_{ij}$ is the
angle between objects $i$ and $j$ and $E_i$ and $E_j$ are the energies
of the objects $i$ and $j$.  The minimum $k_{T\,ij}^2$ was found and
if
\[y_{ij}= \frac{k_{T\,ij}^2}{M_X^2} < y_{\rm{cut}}\] objects $i$ and
$j$ were merged. The merging of the 4-vectors was done using the
recombination ``$E$-scheme'', with simple 4-vector addition, which is
the only Lorentz invariant scheme~\cite{Blazey:2000qt}. It causes the
cluster objects to acquire mass and the total invariant mass, $M_X$,
coincides with the invariant mass of the jet system.  The clustering
procedure was repeated until all $y_{ij}$ values exceeded a given
threshold, $y_{\rm{cut}}$, and all the remaining objects were then labelled
as jets. Applied in the centre-of-mass rest frame, this algorithm
produces at least two jets in every event. The same jet-search
procedure was applied to the final-state hadrons for simulated events.

\begin{figure}
\centering
\includegraphics[width=1\textwidth]{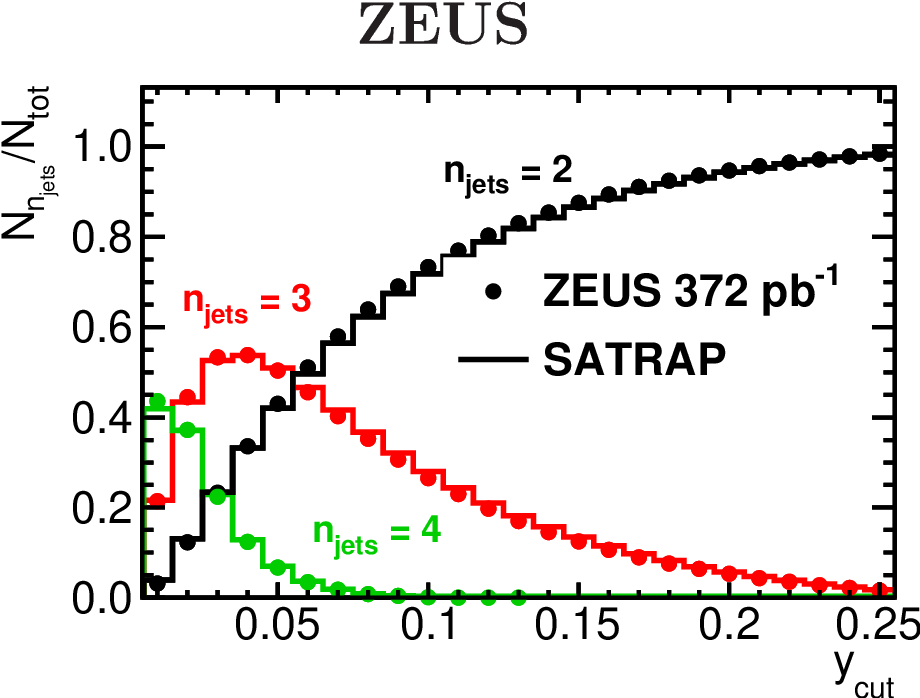}
\caption[]{The probability of finding two, three and four jets in the
  final state as a function of the $y_{cut}$ parameter (see text).
  Data are shown as full dots. Statistical errors are smaller than the
  dot size. Predictions of SATRAP are shown as histograms. The
  distributions are not corrected for detector effects.}
\label{fig_jets_ycut}
\end{figure}

\myFigure{}~\ref{fig_jets_ycut} shows the measured fractions for 2, 3
and 4 jets in the event as a function of the jet resolution parameter,
$y_{\rm{cut}}$~\cite{Catani:1991hj}, in the region
$0.01<y_{\rm{cut}}<0.25$.  The rate of dijet reconstruction varies
from 70\% at $y_{\rm{cut}}=0.1$ to 90\% at $y_{\rm{cut}}=0.2$. The
measured jet fractions were compared to jet fractions predicted by
SATRAP after reweighting of kinematic variables as described in
Section~\ref{sec:mc-vs-data}. SATRAP provides a good description of
the measurement. Jets were reconstructed with a resolution parameter
fixed to $y_{\rm{cut}} = 0.15$. Events with exactly two reconstructed
jets were selected.

Finally, a lower limit of the jet transverse momenta in the
centre-of-mass frame was required, $\ptjet > \unit{2}{\GeV}$. This
value was chosen as a compromise between having a value of $\ptjet$
large enough so that perturbative calculations are still valid and on
the other hand small enough so that a good statistical accuracy can be
still obtained.

\section{Comparison between data and Monte Carlo}
\label{sec:mc-vs-data}

Data and Monte Carlo predictions for several kinematic and jet
variables were compared at the detector level. The MC event
distributions which had been generated with SATRAP were reweighted in
a multidimensional space with respect to: inelasticity
$y$, jet pseudorapidity $\eta_{\rm jet}$, $\ptjet, M_X, Q^2, \beta$
and $x_{\rm I\!P}$. In addition, the prediction of $q \bar q$
production from Bartels et al.~\cite{pl:b379:239} was used for
reweighting in $\phi$.

The background originating from diffractive dijet photoproduction and
non-diffractive dijet production was estimated from Monte Carlo
simulations as described in Section~\ref{sec:monte-carlo}. The
background from beam-gas interactions and cosmic-ray events was
investigated using data taken with empty proton-beam bunches and
estimated to be negligible.

For each of the distributions presented in this section, all selection
criteria discussed above were applied except the cut on the shown
quantity. The estimated background, normalised to the luminosity of 
this analysis, is also shown.

\begin{figure}
\centering
\includegraphics[width=1\textwidth]{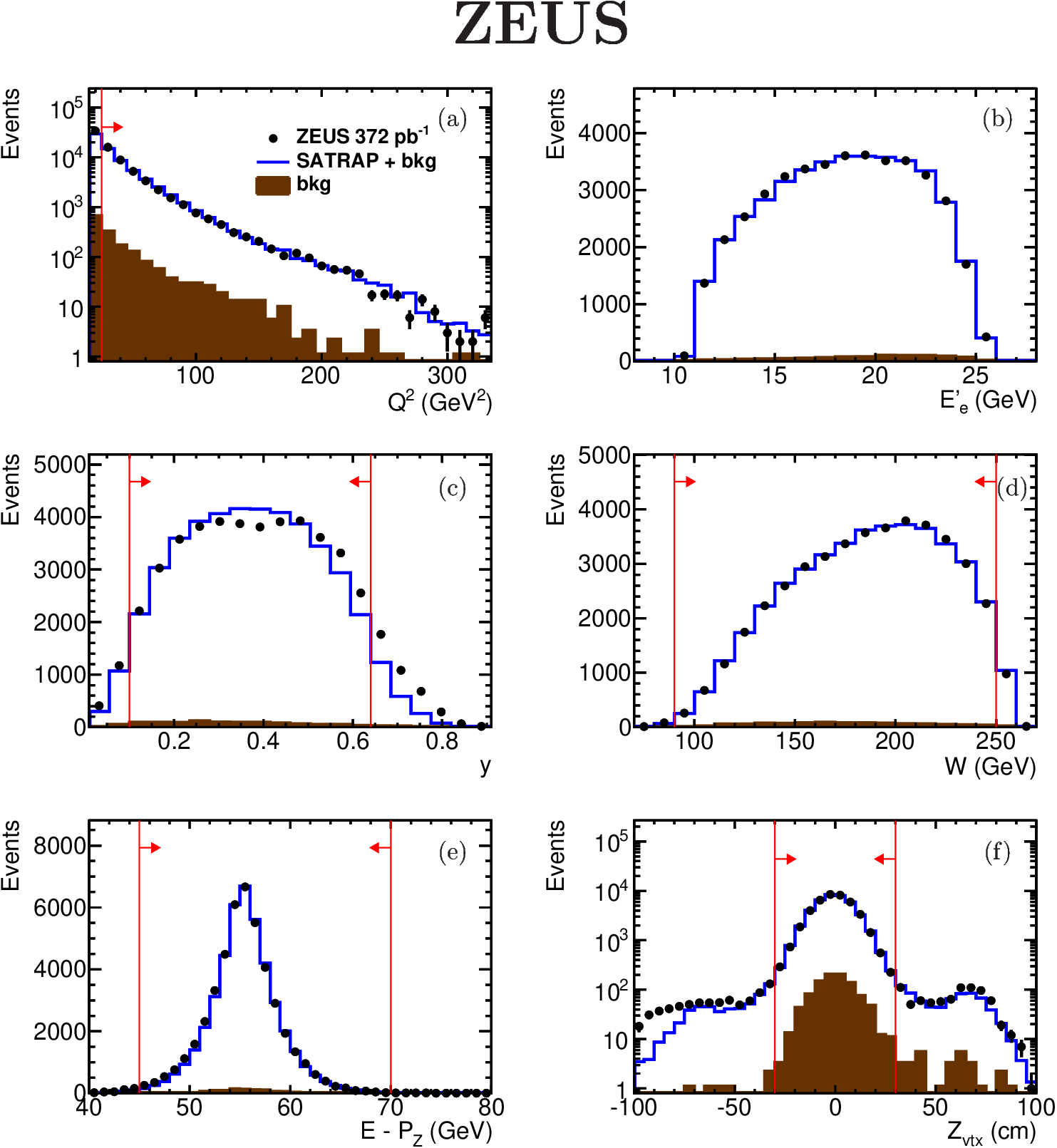}
\caption[]{Comparison between data (dots) and the sum (line) of the
  SATRAP MC and background contributions (shaded), where events with a 
  dissociated proton are not treated as background, for
  kinematic variables: (a) exchanged photon virtuality, $Q^2$, (b)
  scattered electron energy, $E_{e}^\prime$, (c) inelasticity, $y$,
  (d) invariant mass of the $\gamma^*$--$p$ system, $W$, (e) the
  quantity $E-P_Z$ and (f) the $Z$-coordinate of the interaction
  vertex.  The error bars represent statistical errors (generally not
  visible).  The background was normalised to the luminosity and the
  SATRAP MC to the number of events remaining in the data after
  background subtraction. All selection cuts are applied except for the
  cut on the variable shown in each plot.}
\label{fig_var_dis}
\end{figure}

\begin{figure}
\centering
\includegraphics[width=1\textwidth]{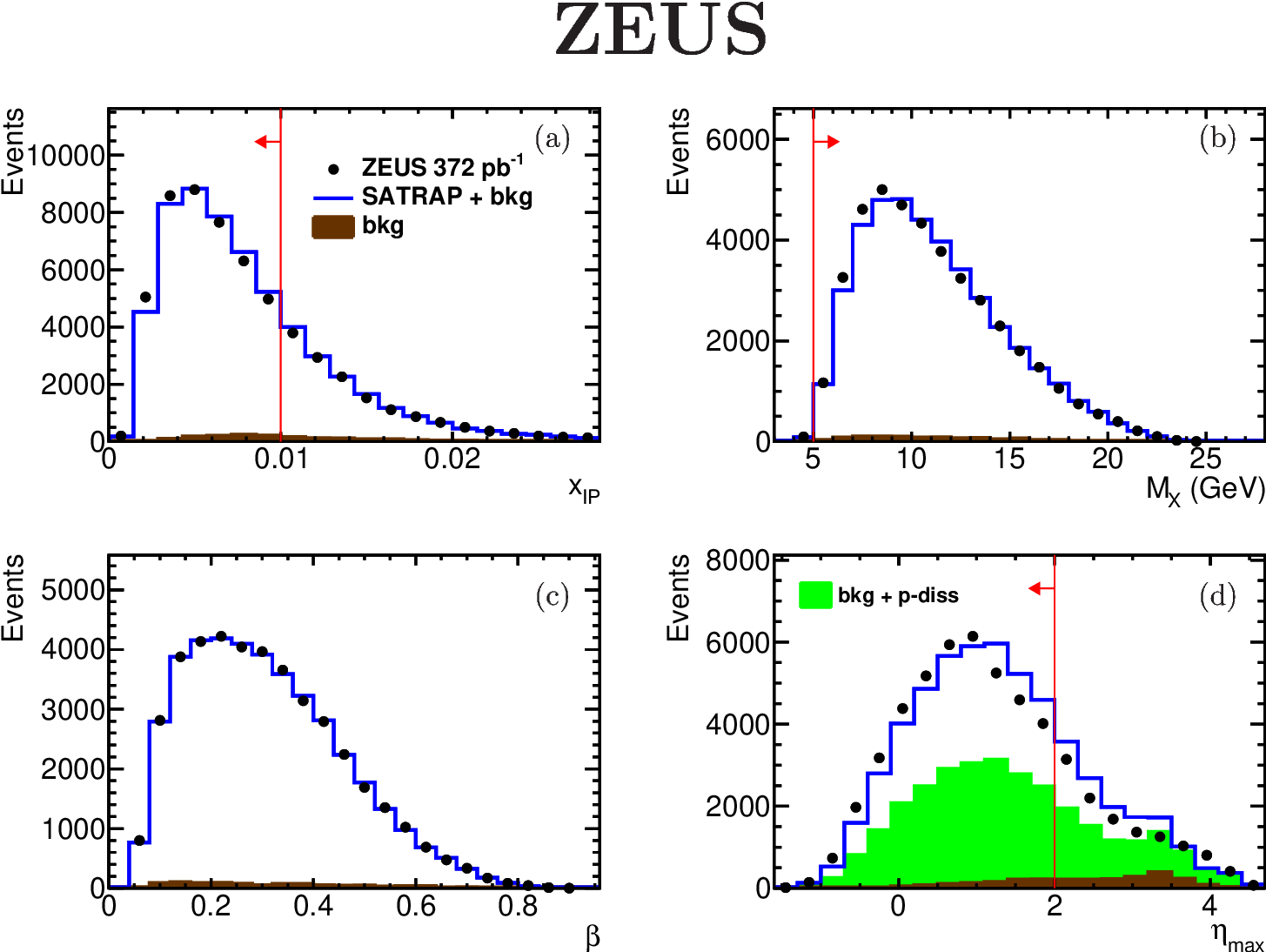}
\caption[]{Comparison between data (dots) and the sum (line) of the SATRAP MC 
and background contributions (dark histogram), where events with a dissociated proton 
were not treated as background, for the kinematic variables of
  the diffractive process: (a) fraction loss in the longitudinal
  momentum of the proton, $x_{\rm I\!P}$, (b) the invariant mass of the
  diffractive system, $M_{\rm X}$, (c) the Bjorken-like variable,
  $\beta$, and (d) the pseudorapidity of the most forward EFO,
  $\eta_{\rm max}$. In (d), the component from events with a
  dissociated proton is also shown (light histogram). Other details as for
  \myFig{}~\ref{fig_var_dis}.}
\label{fig_var_diff}
\end{figure}

\myFigure{}~\ref{fig_var_dis} shows the variables characterising the
DIS events, $Q^2$, $E_e^\prime$, $y$, $W$, $E-P_Z$ and $Z_{\rm{VTX}}$,
while \myFig{}~\ref{fig_var_diff} shows the variables characterising
the diffractive events, $x_{\rm I\!P}$, $M_{\rm X}$, $\beta$ and
$\eta_{\rm max}$. Besides exclusive events, the data contains proton
dissociation, $e + p \rightarrow e + \jetRM{}1 + \jetRM{2} + Y$, for
which the particles stemming from the process of dissociation
disappear undetected in the proton beam hole. Except for
$\eta_{\rm{max}}$, the events with proton dissociation are expected to
yield the same shape of the distributions as the exclusive dijet
events, changing only the normalisation (according to the
factorisation hypothesis, see Section~\ref{sec:monte-carlo}), and are
not shown separately. These events were not treated as a background.
The experimental distributions were compared to the sum of the
background distributions and the SATRAP MC. The background was
normalised to the luminosity and the SATRAP MC to the number of events
remaining in the data after background subtraction.  The
$\eta_{\rm{max}}$ distribution (\myFig{}~\ref{fig_var_diff}(d)) shows
the distribution of events with proton dissociation which was
determined separately as described in Section~\ref{sec:pdiss}. In
\myFig{}~\ref{fig_var_diff}(d), the $\eta_{\rm{max}}$ distribution is
compared to the normalised sum of three contributions including that
of events with proton dissociation. All data distributions, except for
$y$ and $\eta_{\rm max}$, are reasonably well described by the MC
predictions. Most of the difference between data and MC in the $y$
distribution is outside the analysed region ($y > 0.64$). The
incorrect $y$ description at the cut value was taken into account in
the systematic uncertainty which was determined by varying the
cut. The shift in the $\eta_{\rm{max}}$ distribution is accounted for in
the systematic uncertainty of the proton dissociation background (see
Section~\ref{sec:pdiss}).

\begin{figure}
\centering
\includegraphics[width=1\textwidth]{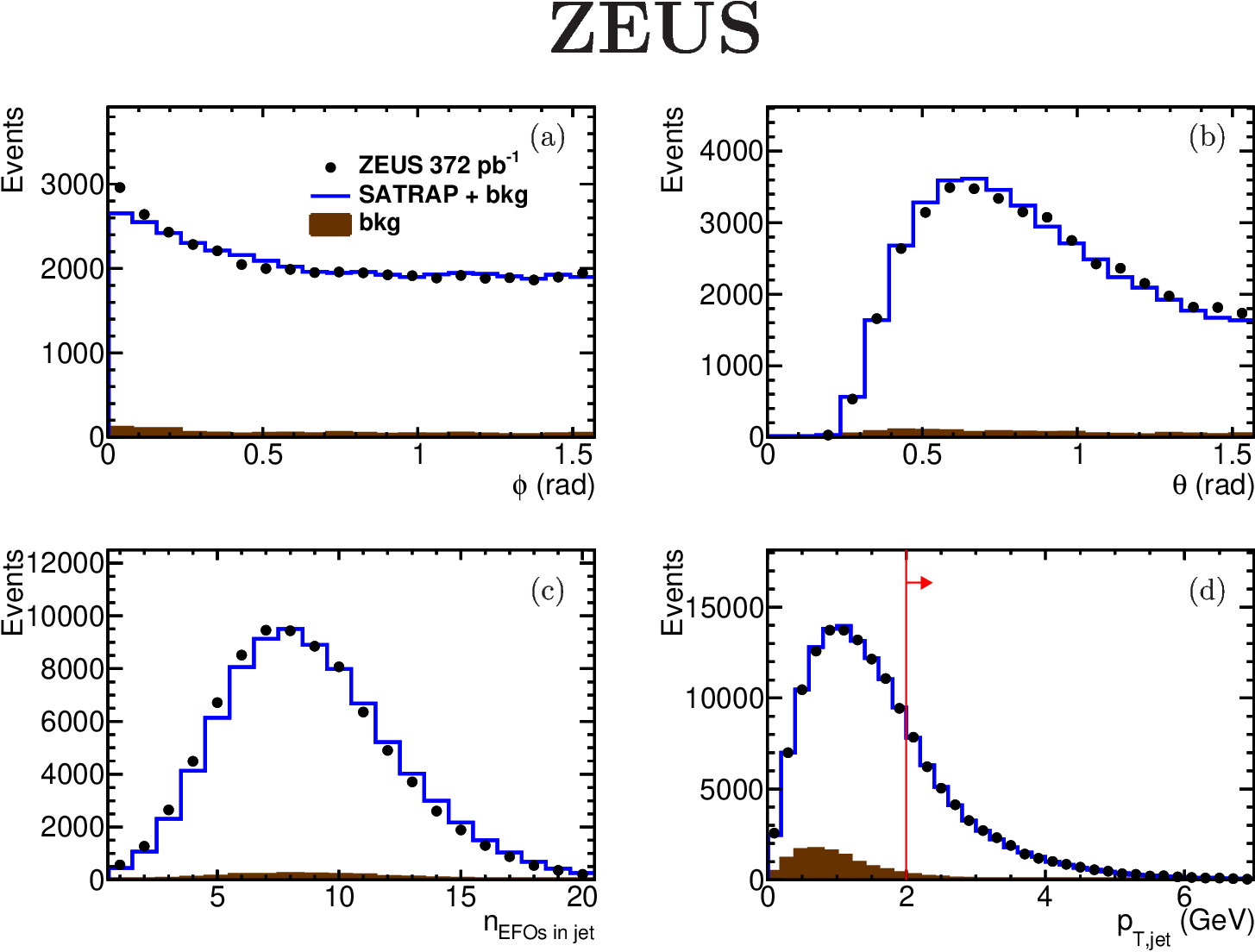}
\caption[]{Comparison between data and MC for variables characterising
  the jet properties in the $\gamma^*$--${\rm I\!P}$ rest frame: (a)
  the azimuthal angle, $\phi$, (b) the polar angle, $\theta$, (c) the
  number of EFOs clustered into a jet and (d) the jet transverse
  momentum, $\ptjet$. Other details as for
  \myFig{}~\ref{fig_var_dis}.}
\label{fig_jets_cms}
\end{figure}

In \myFig{}~\ref{fig_jets_cms}, jet properties in the
$\gamma^*$--${\rm I\!P}$ centre-of-mass system are presented: the
distributions of the jet angles $\theta$ and $\phi$, the number of
EFOs clustered into the jets and the jet transverse momentum $\ptjet$.
All distributions are reasonably well described by the sum of SATRAP
events and the background distribution. The difference between data
and MC for values of $\phi$ close to 0 is not expected to affect
the result of unfolding in this quantity (see Section~\ref{sec:unfolding}).

\begin{figure}
\centering
\includegraphics[width=1\textwidth]{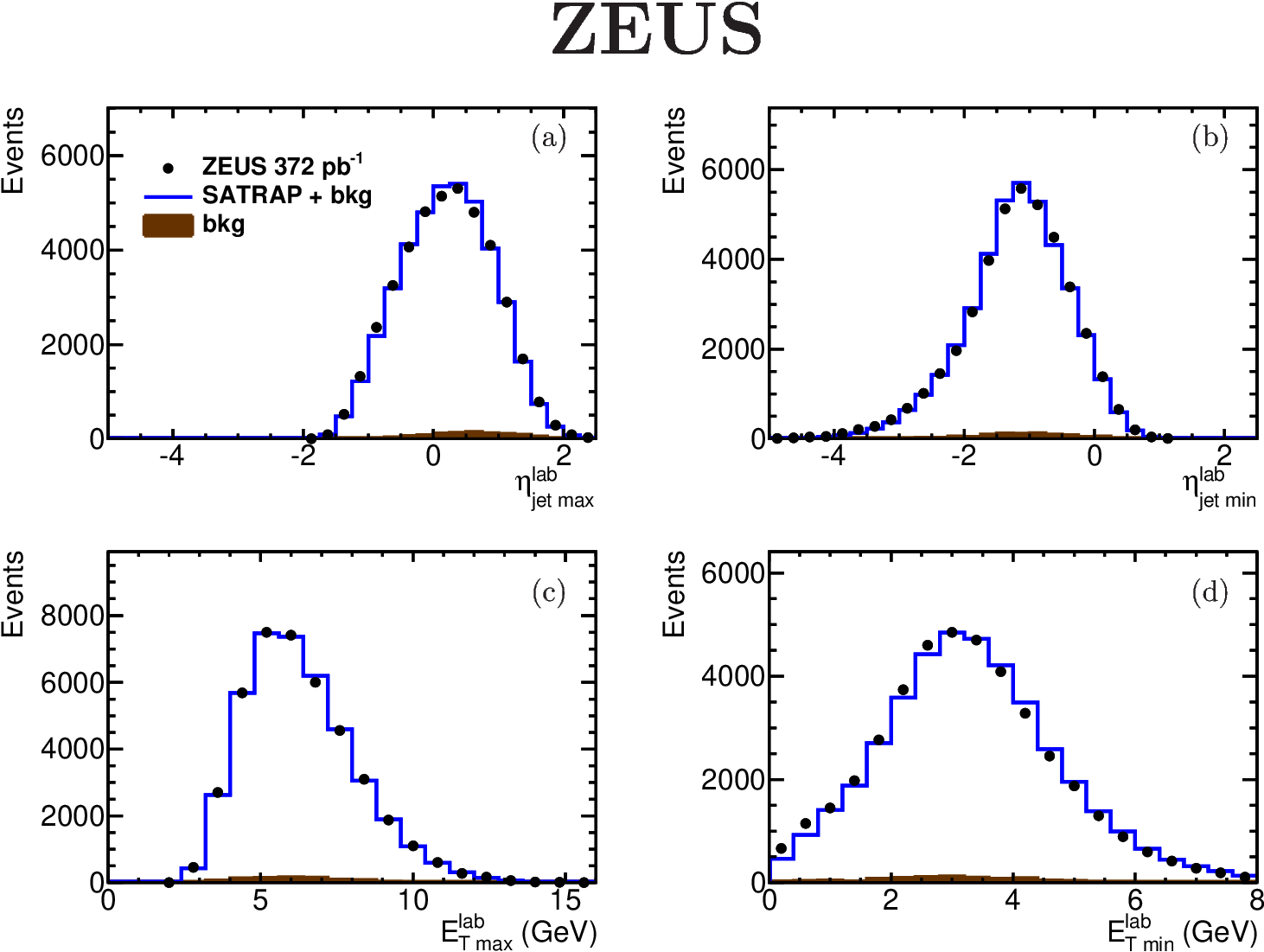}
\caption[]{Comparison between data and MC for variables characterising
  the properties of the higher and the lower energy jets in the LAB
  frame: (a) and (b) the jet pseudorapidity, (b) and (c) the jet
  transverse energy. Other details as for
  \myFig{}~\ref{fig_var_dis}.}
\label{fig_jets_lab}
\end{figure}

Jets reconstructed in the $\gamma^*$--${\rm I\!P}$ rest frame were
transformed back to the laboratory (LAB) system.  In
\myFig{}~\ref{fig_jets_lab} the distributions of the jet
pseudorapidity and the jet transverse energy are shown in the
laboratory system separately for higher- and lower-energy jets. They
are well described by the predicted shape.

The jet algorithm used allows the association of the individual
hadrons with a unique jet on an event-by-event basis. To study the
topology of the jets, the energy flow of particles around the jet axes
was considered in both the centre-of-mass system and the laboratory
system. In this study, $\Delta\eta$ and $\Delta\varphi$ denote the
differences between the jet axis and, respectively, the pseudorapidity
and the azimuthal angle of the EFOs in the event. In
\myFigs{}~\ref{fig_e_flow_cms} the energy flows around the axis of the
reference jet, that is the jet with positive $Z$-component of the
momentum, are shown in the $\gamma^*$--${\rm I\!P}$ centre-of-mass
system. The corresponding distributions in the laboratory system are
presented in \myFigs{}~\ref{fig_e_flow_lab}. It is observed that
energy flows around the reference jet axis are well reproduced by the
SATRAP MC. As expected, the jets are produced back-to-back in the
$\gamma^*$--${\rm I\!P}$ centre-of-mass system, and are quite
broad. However, in the laboratory system, most of their energy is
concentrated within a cone of radius approximately equal to one unit
in the $\eta$--$\varphi$ plane with distance defined as $r =
\sqrt{\Delta\eta^2 + \Delta\varphi^2}$.

The quality of the description of the data by the MC gives confidence
in the use of the MC for unfolding differential cross sections to the
hadron level (see Section~\ref{sec:unfolding}).

\section{Estimate of dijet production with proton dissociation}
\label{sec:pdiss}

The contribution of events with a detected dissociated proton system
is highly suppressed due to the nominal selection cuts applied to the
data, i.e.\ by requiring exactly two jets, $\xPom<0.01$ and
$\eta_{\rm{max}}<2$, and has been considered to be
negligible. However, the contribution of proton-dissociative events,
where the proton-dissociative system escapes undetected, is not
negligible. It was estimated using EPSOFT, after further tuning of the
distribution of the mass of the dissociated proton system, $M_Y$. The
simulation shows that due to the acceptance of the calorimeter,
determined by the detector geometry, a dissociated proton system of
mass smaller than about \unit{6}{\GeV} stays undetected.

In order to estimate the amount of dissociated proton events, a sample
enriched in such events was selected as follows. Kinematic variables
and jets were reconstructed from the EFOs in the range $\eta<2$. All
selection cuts described in Section~\ref{sec:event-select-reconst},
except the $\eta_{\rm{max}}$ cut, were then applied. In order to suppress
non-diffractive contributions to the dijet sample, events with EFOs in
the range $2<\eta<3.5$ were rejected.  The remaining sample of events
with EFOs in the range $\eta>3.5$ consisted almost entirely of
diffractive dijets with a detected dissociated proton system. From the
comparison of the energy sum of all EFOs with $\eta> 3.5$ between data
and simulated events, the following parameterisation of the $M_Y$
distribution was extracted:
\[\frac{ {\rm d}\sigma_{\gamma p\rightarrow \jetRM_1 \jetRM_2 Y}}{{\rm d}M_Y^2} 
\propto {\frac{1}{M_Y^{1.4}}}.\]

The fraction of simulated events with proton dissociation was determined by a fit 
to the distribution of $\eta_{\rm{max}}$  shown in \myFig{}~\ref{fig_var_diff}(d).

The systematic uncertainty of this fraction was estimated in the
following steps:
\begin{itemize}
\item the shape of the $M_Y$ distribution was varied by changing the
  exponent by $\pm 0.6$, because in this way the $\chi^2$ of the
  comparison between data and EPSOFT simulation was raised by 1;

\item the fit of the fraction  
was repeated taking into account a shift of $\eta_{\rm{max}}$ by $+0.1$
according to the observed shift between data and simulated events.
\end{itemize}
Both uncertainties were added in quadrature.

The fraction of events with $\eta_{\rm{max}}<2$ associated to the
proton-dissociative system, which escaped undetected in the beam hole,
was estimated to be $\fpdiss = 45\% \pm 4\% \textrm{(stat.)} \pm 15\%
\textrm{(syst.)}$. No evidence was found that $\fpdiss$ depends on
$\phi$ or $\beta$.  Therefore, in the following sections, the selected data
sample was scaled by a constant factor correcting for
proton-dissociative events.

\section{Unfolding of the hadron-level cross section}
\label{sec:unfolding}

An unfolding method was used to obtain hadron-level differential cross
sections for production of dijets, reconstructed with jet-resolution parameter
$y_{\rm{cut}}=0.15$, as a function of $\beta$ and $\phi$  in the 
following kinematic region: 
\begin{itemize}
\item $Q^2> \unit{25}{\GeV\squared}$; 
\item $90 < W < \unit{250}{\GeV}$;
\item $x_{\rm I\!P} < 0.01$;
\item $M_X > \unit{5}{\GeV}$;
\item $N_{\rm{jets}}=2$;
\item $\ptjet> \unit{2}{\GeV}$.
\end{itemize}

The unfolding was performed by calculating a detector response matrix,
which represents a linear transformation of the hadron-level
two-dimensional distribution of $\phi$--$\ptjet$ or $\beta$--$\ptjet$
to a detector-level distribution. The response matrix was based on the
weighted SATRAP MC simulation. It includes effects of limited detector
and trigger efficiencies, finite detector resolutions, migrations from
outside the phase space and distortions due to QED radiation. The
unfolding procedure was based on the regularised inversion of the
response matrix using Singular Value Decomposition (SVD) as
implemented in the TSVDUnfold package~\cite{Hocker:1995kb}. The
implementation was prepared for one-dimensional problems and the
studied two-dimensional distributions were transformed into
one-dimensional distributions~\cite{Gach:thesis}. The regularisation
parameter was determined according to the procedure suggested by the
authors of the unfolding package.

The used unfolding method takes into account the imperfect hadron level
MC simulation and corrects for it.

\section{Systematic uncertainties}
\label{sec:syst-uncert}

The systematic uncertainties of the cross sections were estimated by
calculating the difference between results obtained with standard and
varied settings for each bin of the unfolded distribution, except for
the uncertainty on $\fpdiss$, which was assumed to give a common
normalisation uncertainty in all the bins.

The sources of systematic uncertainty were divided into two
types. Those originating from detector simulation were investigated by
introducing changes only to MC samples at the detector level, while
the data samples were not altered. The following checks were
performed:
\begin{itemize}
\item the energy scale of the calorimeter objects associated with the
  jet with the highest transverse momentum in the laboratory frame
  was varied by $\pm 5\%$; the corresponding systematic uncertainty
  is in the range of $+2\%,-8\%$;
\item the jet transverse momentum resolution was varied by $\pm
  1\%$, because in this way the $\chi^2$ of the comparison
  of data to MC in the distribution of the jet transverse momentum was
  raised by 1.
\end{itemize}

\begin{figure}
\centering
\includegraphics[width=1\textwidth]{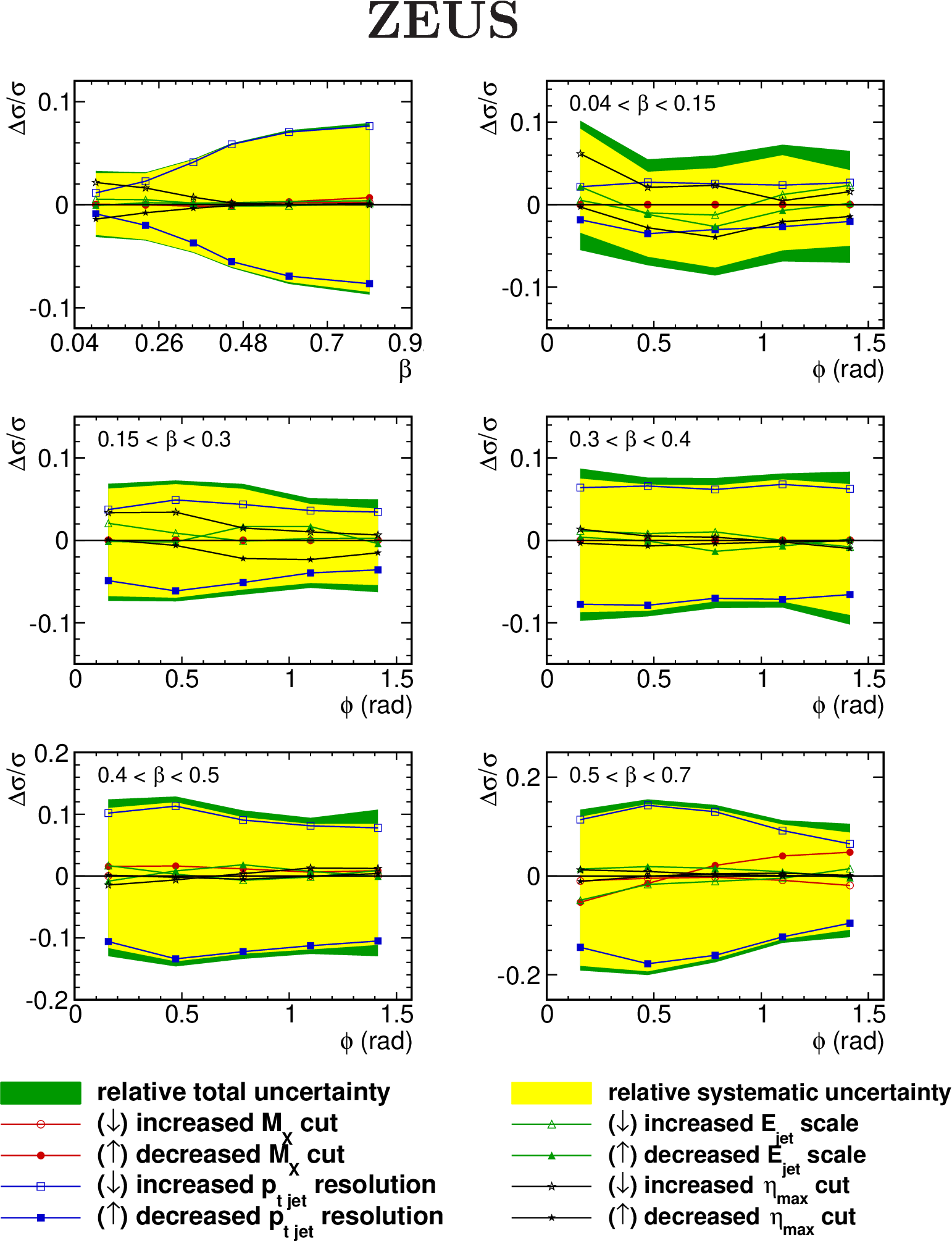}
\vspace{-0.15in}
\caption[]{The most significant sources of systematic uncertainties
  for $\betaCross$ and $\phiCross$ in five $\beta$ ranges.  Total
  systematic uncertainties and total uncertainties (systematic and
  statistical added in quadrature) are shown as shaded and dark-shaded
  bands.}
\label{fig_systematics}
\end{figure}

Systematic effects originating from event-selection cuts were
investigated by varying the criteria used to select events for both
data and simulated events in the following ways:
\begin {itemize} 
\item $Q^2>25 \pm \unit{1.7}{\GeV\squared}$;
\item $90 \pm \unit{7.4}{\GeV} < W < 250 \pm \unit{8.4}{\GeV}$;
\item $0.1 \pm 0.04 < y < 0.64 \pm 0.03$;
\item $ \mid Z_{\textrm{VTX}} \mid < 30 \pm \unit{5}{\centi\metre}$;
\item $ x_{\rm I\!P} < 0.01 \pm 0.001$;
\item $ \etaMax < 2 \pm 0.2$;
\item $M_X > 5 \pm \unit{0.8}{\GeV}$;
\item $E_{\textrm{EFO}} >  0.4 \pm \unit{0.1}{\GeV}$.
\end{itemize}
The uncertainty related to the $M_X$-cut variation is in the range of
$\pm 5\%$. The total uncertainty related to the event-selection cuts
excluding the $M_X$ cut is smaller than $\pm 6\%$. Uncertainties
originating from the most significant sources are presented in
\myFig{}~\ref{fig_systematics}.

Positive and negative uncertainties were separately added in
quadrature. The corresponding total systematic uncertainty is also
shown in \myFig{}~\ref{fig_systematics}. The normalisation uncertainty
of the cross section related to the luminosity (see
Section~\ref{sec-exp}) as well as to $\fpdiss$ is not shown on the
following figures but is included as a separate column in the tables
of cross sections. The total uncertainties of the measured cross
sections are dominated by the systematic component.

\section{Cross sections}
\label{sec:cross-sections}

Cross sections were measured at the hadron level in the kinematic
range described in Section~\ref{sec:unfolding}. Backgrounds from
diffractive photoproduction and non-diffractive dijet production were
subtracted.

In order to calculate the cross sections for exclusive dijet
production, the measured cross sections were scaled by a factor of
$(1-\fpdiss)=0.55$ according to the estimate of the
proton-dissociative background described in Section~\ref{sec:pdiss}.

\begin{figure}
\centering
\includegraphics[width=1\textwidth]{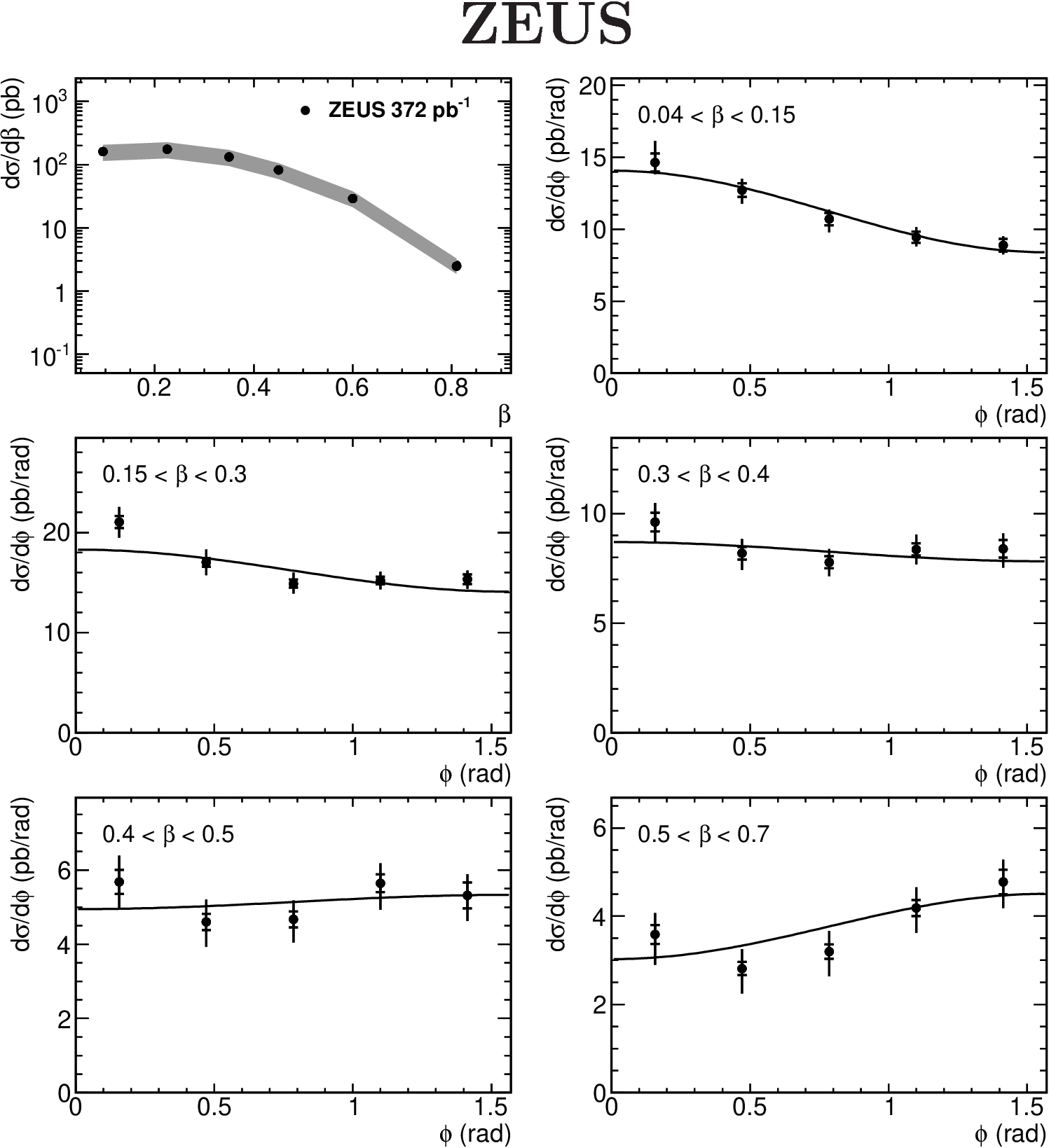}
\vspace{-0.15in}
\caption[]{Differential cross sections for exclusive dijet production:
  $\betaCross$ (in log scale) and $\phiCross$ (in linear scale) in
  five bins of $\beta$. Contributions from proton-dissociative dijet
  production were subtracted. The full line represents the fitted
  function proportional to $1+ A \cosTwoPhi$. Statistical and
  systematic uncertainties were included in the fit. The total error
  bars show statistical and systematic uncertainties added in
  quadrature. The statistical uncertainties were taken from the
  diagonal elements of the covariance matrix. The systematic
  uncertainties do not include the uncertainty of the subtraction of
  the proton-dissociative contribution. This normalisation uncertainty
  is shown as a grey band only in the $\betaCross$ distribution.}
\label{fig_kinematics_noTheory}
\end{figure}

The values of the cross-sections $\betaCross$ and $\phiCross$ in five
bins of $\beta$ are given in Tables~\ref{tab:beta} and~\ref{tab:phi}
and shown in \myFig{}~\ref{fig_kinematics_noTheory}. The statistical
uncertainties presented in the figures correspond to the diagonal
elements of the covariance matrices, which are available in electronic
format~\cite{misc:covMat}. The $\betaCross$ distribution is, due to
the kinematics, restricted to the range $0.04<\beta< 0.92$.  The
$\phiCross$ distribution is shown in five bins of $\beta$ in the range
$0.04<\beta<0.7$. The cut at 0.7 excludes a region with a low number
of events.

\begin{figure}
\centering
\includegraphics[width=0.95\textwidth]{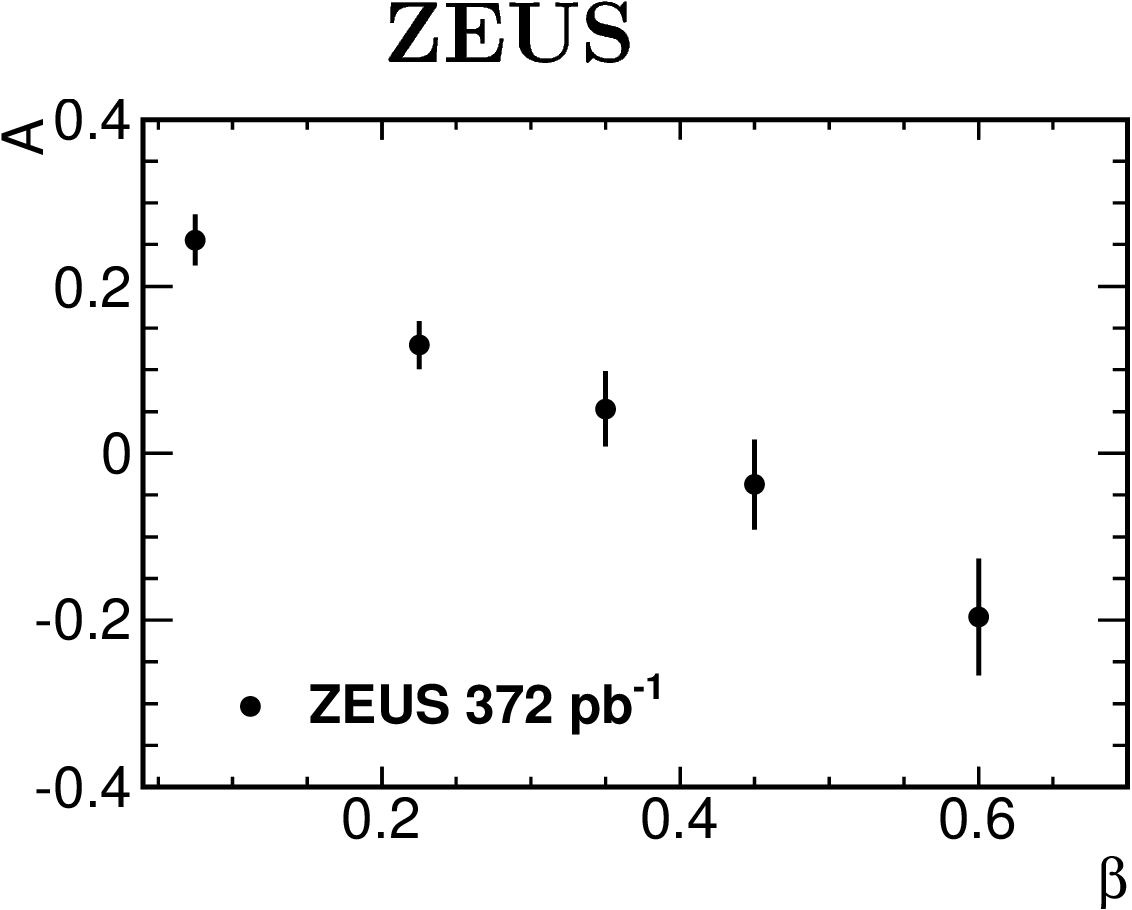}
\vspace{-0.15in}
\caption[]{The shape parameter $A$ as a function of $\beta$ resulting
  from the fits to $\phiCross$ with a function proportional to $1+A
  \cosTwoPhi$. The statistical and systematic uncertainties were included
  in the fit.}
\label{fig_aNoTheory}
\end{figure}

The $\phi$ distributions show a significant feature: when going from
small to large values of $\beta$, the shape varies and the slope of
the angular distribution changes sign. The variation of the shape was
quantified by fitting a function to the $\phi$ distributions including
the full statistical covariance matrix and the systematic
uncertainties, the latter by using the profile
method~\cite{Blobel:2003wa}. The fitted function is predicted by
theoretical calculations (see Section~\ref{sec-int}) to be proportional to
($1+A \cosTwoPhi$). The data are well described by the fitted
function. The resulting values of $A$ are shown in
Table~\ref{tab:AvsBeta} and \myFig{}~\ref{fig_aNoTheory}. The
parameter $A$ decreases with increasing $\beta$ and changes sign
around $\beta=0.4$.

\section{Comparison with model predictions}

\begin{figure}
\centering
\includegraphics[width=0.6\textwidth]{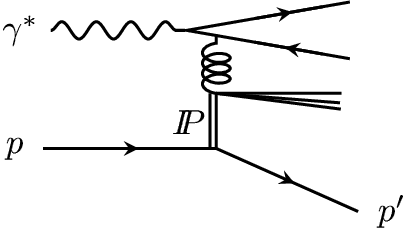}
\vspace{-0.15in}
\caption[]{Diagram of diffractive boson--gluon fusion in the
  Resolved-Pomeron model.}
\label{fig_bgf}
\end{figure}

The differential cross sections were compared to MC predictions for
the Resolved-Pomeron model and the Two-Gluon-Exchange model.
In the Resolved-Pomeron model~\cite{pl:b152:256}, the diffractive
scattering is factorised into a Pomeron flux from the proton and the
hard interaction between the virtual photon and a constituent parton
of the Pomeron.  An example of such a process is shown in
\myFig{}~\ref{fig_bgf}, where a $q \bar q$ pair is produced by a
boson--gluon fusion (BGF) process associated with the emission of a
Pomeron remnant. This model requires the proton diffractive gluon
density as an input for the calculation of the cross section. The
predictions considered in this article are based on the
parameterisation of the diffractive gluon density obtained from fits
(H1 2006 fits A and B) to H1 inclusive diffractive
data~\cite{Aktas:2006hy}. The shape of the $\phi$ distribution is
essentially identical in all models based on the BGF process,
including both the Resolved-Pomeron and the Soft Colour Interactions
(SCI) model~\cite{pl:b366:371}.

\begin{figure}
\centering
\includegraphics[width=0.6\textwidth]{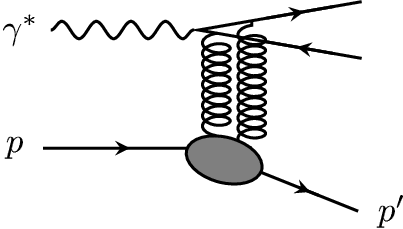}
\vspace{-0.15in}
\caption[]{Example diagram of $q\bar{q}$ production in the Two-Gluon-Exchange model.}
\label{fig_two_gluon_qq}
\end{figure}

\begin{figure}
\centering
\includegraphics[width=0.6\textwidth]{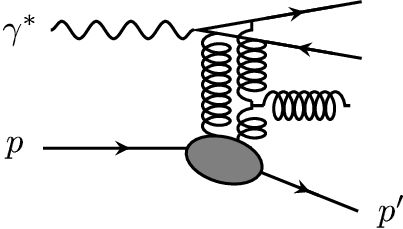}
\vspace{-0.15in}
\caption[]{Example diagram of $q\bar{q}g$ production in the Two-Gluon-Exchange model.}
\label{fig_two_gluon_qqg}
\end{figure}

In the Two-Gluon-Exchange model~\cite{pl:b379:239, pl:b386:389,
  epj:c11:111, Braun:2005rg}, the diffractive production of a $q \bar
q$ pair is due to the exchange of a two-gluon colour-singlet
state. The process is schematically shown in
\myFig{}~\ref{fig_two_gluon_qq}. The $q \bar q$ pair hadronises into a
dijet final state.  For large diffractive masses, i.e.\ at low values
of $\beta$, the cross section for the production of a $q \bar q$ pair
with an extra gluon is larger than that of the $q \bar q$
production. The diagram of this process is shown in
\myFig{}~\ref{fig_two_gluon_qqg}.  The $q\bar{q}g$ final state also
contributes to the dijet event sample if two of the partons are not
resolved by the jet algorithm (see Section~\ref{sec:jet-selection}).

The $q\bar{q}$ pair production was calculated to second order in QCD,
using the running strong-interaction coupling constant $\alpha_s(\mu)$
with the scale $\mu=p_{T}\sqrt{1+Q^2/M_X^2}$~\cite{pl:b379:239,
  pl:b386:389}, where $p_{T}$ denotes the transverse momentum of the
quarks in the $\gamma^*$--${\rm I\!P}$ rest frame with respect to the
virtual photon momentum and $M_X$ is the invariant mass of the
diffractive system. The prediction was calculated with a cut on the
transverse momentum of the quarks, $p_T > \unit{1}{\GeV}$.  The cross
section is proportional to the square of the gluon density of the
proton, $g(x_{\rm I\!P},\mu^2)$
\[d\sigma \propto \left[ \frac{\alpha_s(\mu)}{p_{T}^2}x_{\rm I\!P}g(x_{\rm I\!P},\mu^2) \right]^2.\]
 
The cross section for the $q\bar{q}g$ final state was calculated
taking into account that it is proportional to the square of the gluon
density of the proton, $g(x_{\rm I\!P},\hat{k}_T^2)$, at a scale
$\hat{k}_T^2$, which effectively involves the transverse momenta of all
three partons. For the calculation of the cross
section, a fixed value of
$\alpha_s=0.25$~\cite{epj:c11:111} and the GRV~\cite{zfp:c67:433}
parameterisation of the gluon density were used and the same cut was
applied on the transverse momentum of all partons:
$p_{T,\mathrm{parton}} > \ptcut$ with the value adjusted to the data
(see Section~\ref{sec:qqFrac}). In contrast to $q\bar{q}$ production, the
exclusive dijet cross section calculated for the $q\bar{q}g$ final
state is sensitive to the parton-level
cut $\ptcut$. This is a consequence of the fact that two of the partons
form a single jet.

\subsection{Contribution of the $\boldsymbol{q \bar q}$ dijet
  component in the prediction of the Two-Gluon-Exchange model}
\label{sec:qqFrac}

\begin{figure}
\centering
\includegraphics[width=1\textwidth]{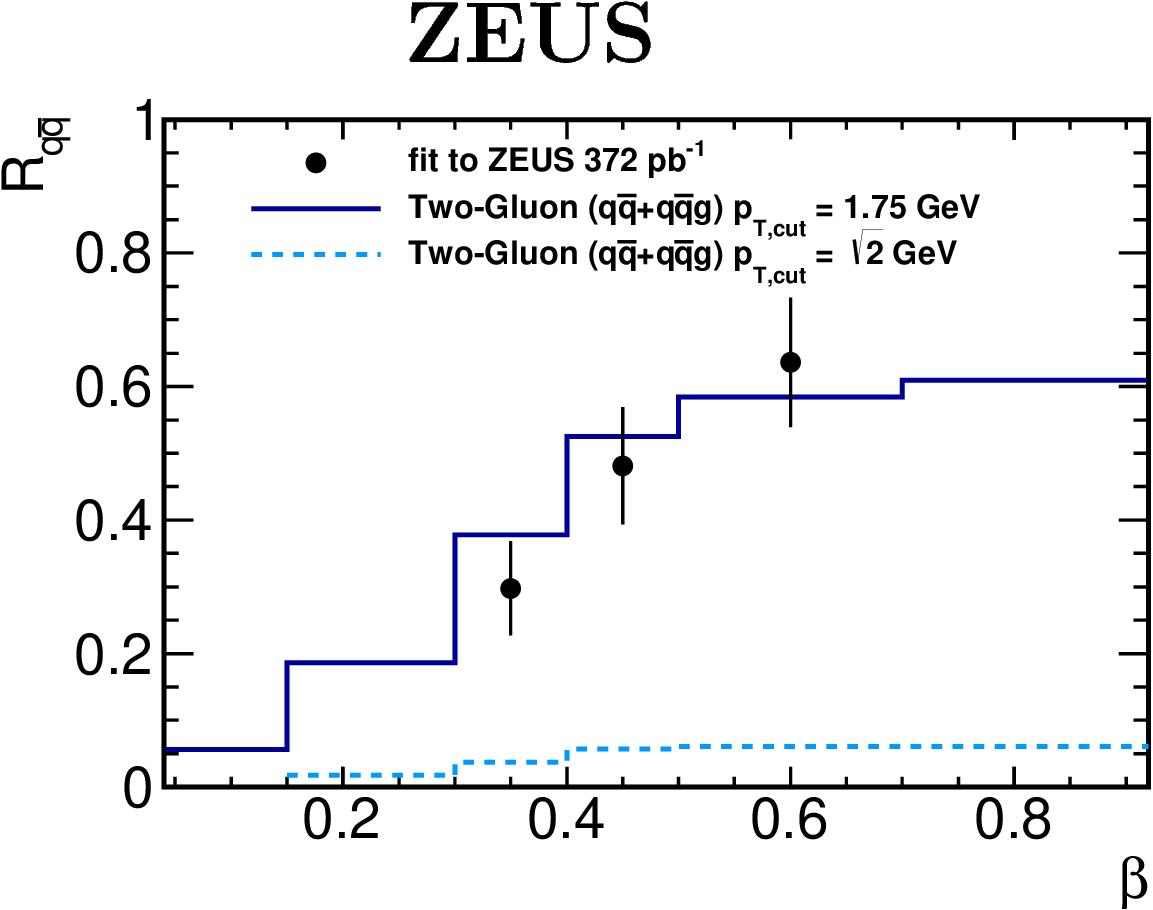}
\vspace{-0.15in}
\caption[] 
{The $R_{q \bar q} =
\sigma(q\bar{q})/(\sigma(q\bar{q})+\sigma(q\bar{q}g))$, determined in a
fit of the predicted shapes to the measured $\phi$ distributions given
in \myFig{}~\ref{fig_kinematics_noTheory}. The fit takes into account the
full covariance matrix. The predicted ratio is shown for two choices
of $\ptcut$: for the \unit{$\sqrt{2}$}{\GeV} used for the published
calculations~\cite{epj:c11:111} and for 
$\unit{1.75}{\GeV}$, determined in a fit.}
\label{fig_fractions}
\end{figure}

In the Two-Gluon-Exchange model, the $\phi$ distribution predicted for
$q \bar q$ and $q\bar{q} g$ have different shapes. This allows the
ratio $R_{q \bar q} = \sigma(q \bar {q})/\sigma(q \bar {q} + q \bar
{q} g)$ to be determined by studying the measured $\phi$
distributions. The results are shown in
\myFig{}~\ref{fig_fractions}. The ratio was measured only in the
region of $\beta \in (0.3, 0.7)$ since elsewhere the uncertainty
estimation is unreliable due to the measured value being too close to
0 or 1. The ratio $R_{q \bar q}$ predicted by the model depends on the
parton transverse-momentum cut applied. The $\ptcut$ value of
\unit{$\sqrt{2}$}{\GeV} used in the original
calculation~\cite{epj:c11:111} significantly underestimates the
ratio. A scan of the parton transverse-momentum cut showed that the
measured ratio can be well described throughout the considered range
with $\ptcut = \unit{1.75}{\GeV}$. Both this value of $\ptcut$ and the
original value were used for calculating the Two-Gluon-Exchange model
predictions.

\subsection{Differential cross-section $\boldsymbol{\betaCross}$}

\begin{figure}
\centering
\includegraphics[width=1\textwidth]{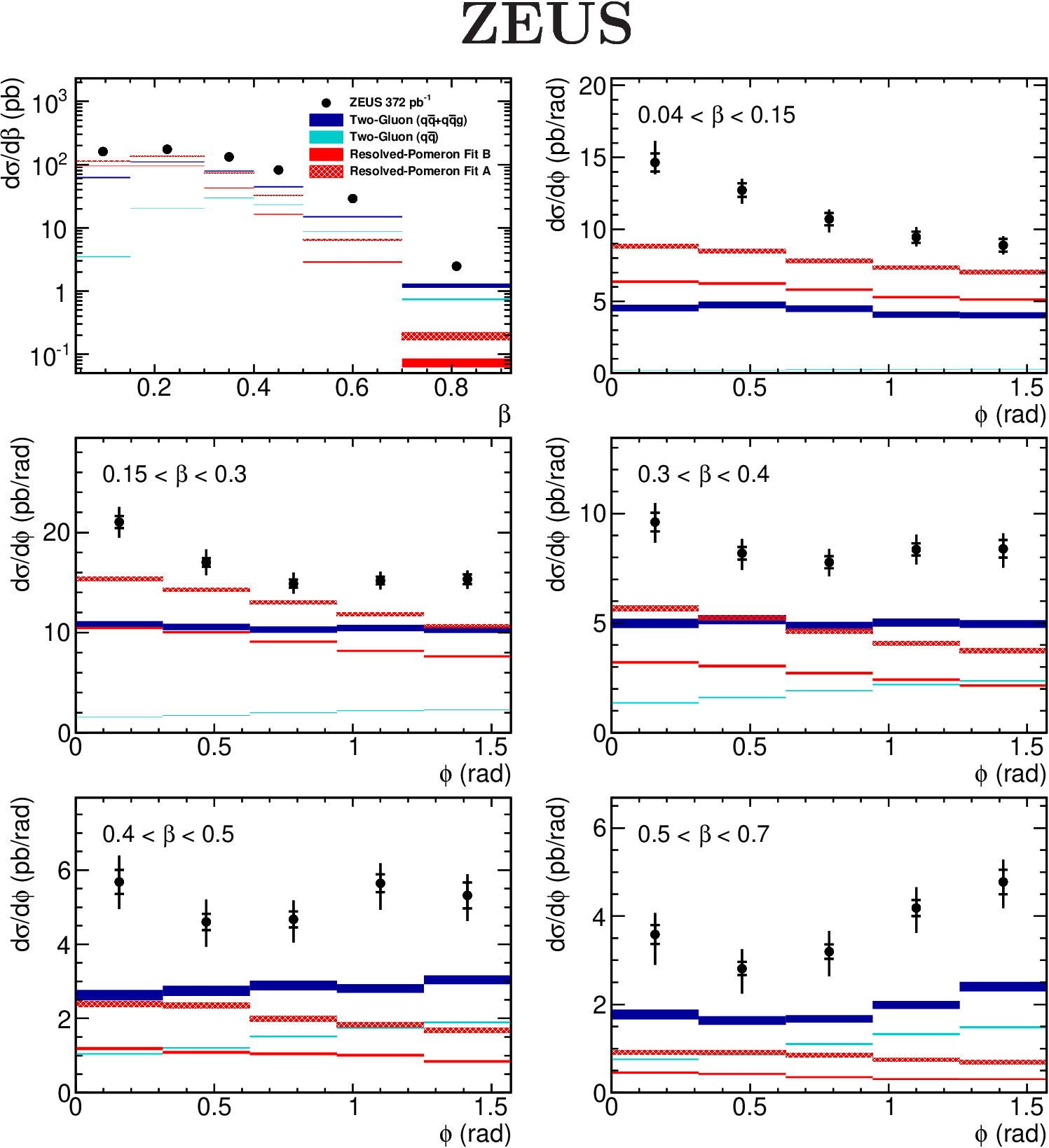}
\vspace{-0.15in}
\caption[]{Differential cross sections as in
  \myFig{}~\ref{fig_kinematics_noTheory} in comparison to model
  predictions $\betaCross$ (in log scale) and $\phiCross$ in bins of
  $\beta$ (in linear scale). Contributions from proton-dissociative
  dijet production were subtracted. The systematic uncertainties do
  not include the uncertainty due to the subtraction. The
  Two-Gluon-Exchange model is presented with $\ptcut =
  \unit{1.75}{\GeV}$. The bands on theoretical expectations represent
  statistical uncertainties only.}
\label{Final}
\end{figure}

The cross-section $\betaCross$ is shown in \myFig{}~\ref{Final} together
with the predictions from both models. The prediction of the
Resolved-Pomeron model decreases with increasing $\beta$ faster than
the measured cross section, for both fit A and fit B. The difference
between data and prediction is less pronounced for fit A than for fit
B, which is consistent with the observation that the ratio of gluon
densities increases with increasing
$\beta$~\cite{Aktas:2006hy}. Predictions and data differ by a factor
of two for small values of $\beta$ and about ten for large values.

The Two-Gluon-Exchange model prediction, which includes $q \bar q $
and $q \bar q g$, describes the shape of the measured $\beta$
distribution reasonably well. The predicted integrated cross section
is $\sigma = \unit{38}{\pico\barn}$, while the measured cross section
is $\sigma = \unit{72}{\pico\barn}$ with a normalisation uncertainty
originating from the proton-dissociation background of $u(\fpdiss) /
(1 - \fpdiss)= 27\%$, where $u(\fpdiss)$ is the uncertainty
in the fraction of events with a dissociated proton. Although the difference
between the predicted and measured cross section is not significant, it
could indicate that the NLO corrections are large or the cross-section
enhancement arising from the evolution of the off-diagonal gluon
distribution is significant~\cite{GolecBiernat:1998vf}. The
prediction based on $q\bar q$ production alone fails to describe the
shape of the distribution at low values of $\beta$ but is almost
sufficient to describe it at large $\beta$, where the $q \bar q g$
component is less important.

\subsection{Differential cross-section $\boldsymbol{\phiCross}$}

\begin{figure}
\centering
\includegraphics[width=1\textwidth]{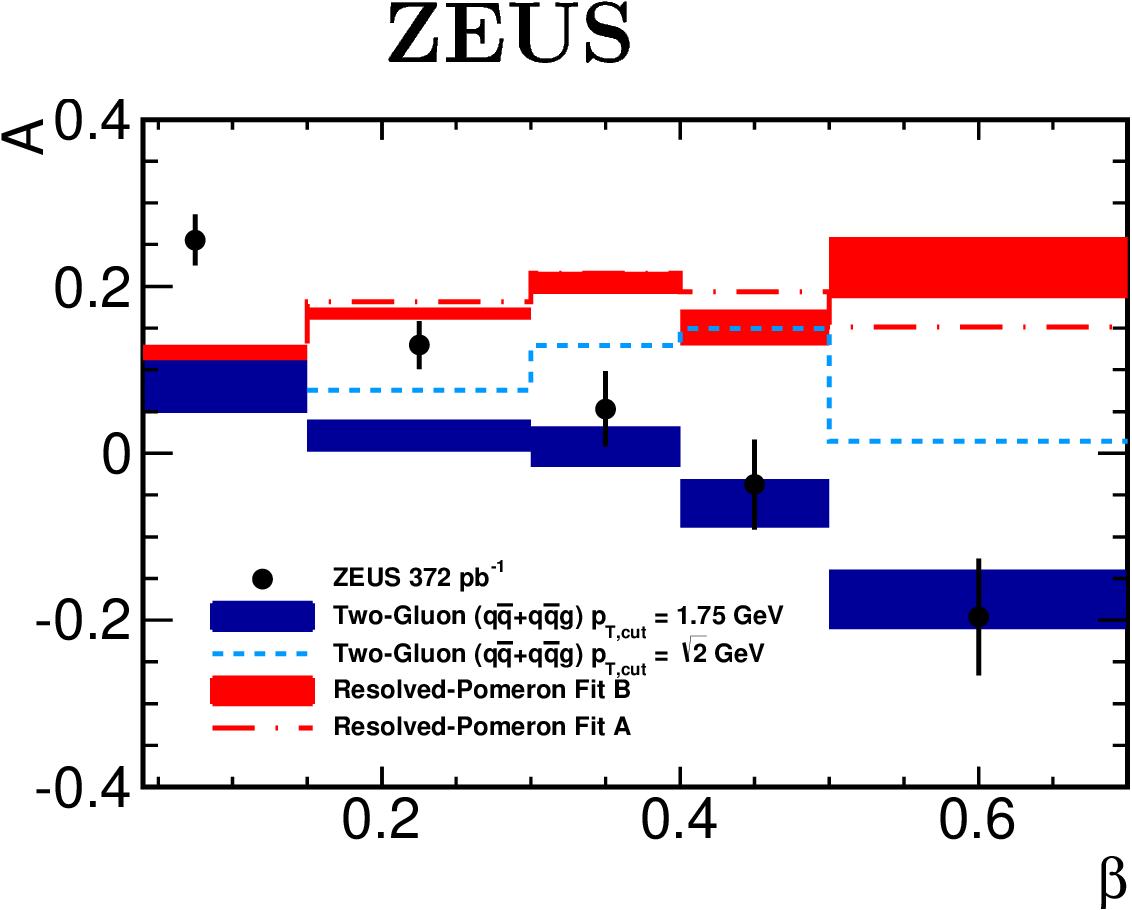}
\vspace{-0.15in}
\caption[]{The shape parameter $A$ as a function of $\beta$ in
  comparison to the values of $A$ obtained from distributions
  predicted by the Resolved-Pomeron model and the Two-Gluon-Exchange
  model. The bands on BGF Fit B and two-gluon $\ptcut=
  \unit{1.75}{\GeV}$ represent statistical uncertainties.}
\label{fig_a}
\end{figure}

The cross-sections $\phiCross$ are shown in \myFig{}~\ref{Final} in
five different $\beta$ ranges together with the predictions of both
models. The comparison of the shapes has been quantified by
calculating the slope parameter $A$. The results are shown in
\myFig{}~\ref{fig_a}. The Resolved-Pomeron model predicts an almost
constant, positive value of $A$ in the whole $\beta$ range. The
Two-Gluon-Exchange model ($q \bar{q} + q \bar{q} g$) predicts a value
of $A$ which varies from positive to negative. In contrast to the
Resolved-Pomeron model, the Two-Gluon-Exchange model agrees
quantitatively with the data in the range $0.3 < \beta < 0.7$. The
prediction based on $q \bar q$ production alone describes the shape of
the distributions at large $\beta$, where the $q \bar{q} g$ component
is less important.

\section{Summary}

The first measurement of diffractive production of exclusive dijets in
deep inelastic scattering, $\gamma^* + p \rightarrow \jetRM{}1 +
\jetRM{}2 + p$, was presented.  The differential cross-sections
$\betaCross$ and $\phiCross$ in bins
of $\beta$ were measured in the kinematic range: $Q^2 >
\unit{25}{\GeV\squared}$, $90 < W < \unit{250}{\GeV}$, $M_X >
\unit{5}{\GeV}$, $x_{\rm I\!P} < 0.01$ and $\ptjet > \unit{2}{\GeV}$
using an integrated luminosity of $\unit{372}{\invpb}$.

The measured absolute cross sections are larger than those predicted
by both the Resolved-Pomeron and the Two-Gluon-Exchange models. The
difference between the data and the Resolved-Pomeron model at $\beta >
0.4$ is significant.  The Two-Gluon-Exchange model predictions agree
with the data within the experimental uncertainty and are themselves
subject to possible large theoretical uncertainties. The shape of the
$\phi$ distributions was parameterised with the function $1 + A
\cosTwoPhi$, as motivated by theory. The Two-Gluon-Exchange model predicts reasonably well the
measured value of $A$ as a function of $\beta$, whereas the
Resolved-Pomeron model exhibits a different trend.

\section*{Acknowledgements}
\label{sec-ack}
We appreciate the contributions to the construction, maintenance and
operation of the ZEUS detector of many people who are not listed as
authors. The HERA machine group and the DESY computing staff are
especially acknowlegded for their success in providing excellent
operation of the collider and the data-analysis environment. We thank
the DESY directorate for their strong support and encouragement.  We
thank J. Bartels and H. Jung for their help with the theoretical
predictions.

 
\vfill\eject

\clearpage
{\raggedright
\providecommand{\etal}{et al.\xspace}
\providecommand{\coll}{Collab.\xspace}
\catcode`\@=11
\def\@bibitem#1{%
\ifmc@bstsupport
  \mc@iftail{#1}%
    {;\newline\ignorespaces}%
    {\ifmc@first\else.\fi\orig@bibitem{#1}}
  \mc@firstfalse
\else
  \mc@iftail{#1}%
    {\ignorespaces}%
    {\orig@bibitem{#1}}%
\fi}%
\catcode`\@=12

}

\clearpage

\begin{table}
  \centering
  \caption{ Differential cross-section $\betaCross$ in the kinematic range:
    $Q^2 > \unit{25}{\GeV\squared}$, $90 < W < \unit{250}{\GeV}$,  $x_{\rm
      I\!P} < 0.01$, $M_X > \unit{5}{\GeV}$ and  $\ptjet > \unit{2}{\GeV}$.
    The statistical uncertainties are given by the diagonal part of the covariance matrix.
    Systematic uncertainties are explained in the text. 
    The contribution from proton dissociation was subtracted. The uncertainty of the subtraction determines the uncertainty of the normalisation 
    also given in the table.  
    \vspace*{1cm}}
  \centering
{\renewcommand{\arraystretch}{1.35}
\begin{tabular}{r c r r@{$\pm$}p{1.6cm} l}
\noalign{\hrule height 1.25pt} 
\multicolumn{3}{c}{ $\beta$ } & \multicolumn{3}{c}{ $\betaCross $ (pb)} \\ 
\hline 
0.04 & -- & 0.15 & 159.7 & 1.8(stat.) & $^{+6.0}_{-4.7}$(sys.)$\pm 45.2$(norm.) \\ 
0.15 & -- & 0.3 & 175.1 & 1.3(stat.) & $^{+6.7}_{-6.0}$(sys.)$\pm 49.6$(norm.) \\ 
0.3 & -- & 0.4 & 132.3 & 1.2(stat.) & $^{+6.0}_{-6.1}$(sys.)$\pm 37.5$(norm.) \\ 
0.4 & -- & 0.5 & 82.1 & 0.9(stat.) & $^{+5.0}_{-5.0}$(sys.)$\pm 23.3$(norm.) \\ 
0.5 & -- & 0.7 & 29.0 & 0.5(stat.) & $^{+2.1}_{-2.2}$(sys.)$\pm 8.3$(norm.) \\ 
0.7 & -- & 0.92 & 2.47 & 0.06(stat.) & $^{+0.20}_{-0.21}$(sys.)$\pm 0.70$(norm.) \\ 
\noalign{\hrule height 1.25pt} 
\end{tabular}
}
  \label{tab:beta}
\end{table}

\begin{table}
  \centering
  \caption{Differential cross-section $\phiCross$  in the kinematic range:
    $Q^2 > \unit{25}{\GeV\squared}$, $90 < W < \unit{250}{\GeV}$,  $x_{\rm
      I\!P} < 0.01$, $M_X > \unit{5}{\GeV}$ and  $\ptjet > \unit{2}{\GeV}$. 
    Statistical uncertainties are given by the diagonal elements of the covariance matrix. Systematic uncertainties 
    are explained in the text. The contribution from
    proton dissociation was  subtracted. The  uncertainty of the subtraction determines the 
    uncertainty of the normalisation given in the table.}
  \centering
{\renewcommand{\arraystretch}{1.35}
\begin{tabular}{l c l r@{$\pm$}p{1.5cm} l}
\noalign{\hrule height 1.25pt} 
\hline 
\multicolumn{3}{c}{ $\phi $ (rad)} & \multicolumn{3}{c}{ $\phiCross$ (pb/rad)} \\ 
\hline 
\hline 
\multicolumn{6}{c}{$0.04 < \beta < 0.15$} \\ 
\hline 
0 & -- & 0.314 & 14.64 & 0.64(stat.) & $^{+1.37}_{-0.50}$(sys.)$\pm 4.15$(norm.) \\ 
0.314 & -- & 0.628 & 12.73 & 0.49(stat.) & $^{+0.62}_{-0.81}$(sys.)$\pm 3.60$(norm.) \\ 
0.628 & -- & 0.942 & 10.71 & 0.43(stat.) & $^{+0.51}_{-0.82}$(sys.)$\pm 3.03$(norm.) \\ 
0.942 & -- & 1.26 & 9.46 & 0.39(stat.) & $^{+0.58}_{-0.53}$(sys.)$\pm 2.68$(norm.) \\ 
1.26 & -- & 1.57 & 8.89 & 0.45(stat.) & $^{+0.45}_{-0.45}$(sys.)$\pm 2.52$(norm.) \\ 
\hline 
\multicolumn{6}{c}{$0.15 < \beta < 0.3$} \\ 
\hline 
0 & -- & 0.314 & 21.03 & 0.60(stat.) & $^{+1.38}_{-1.43}$(sys.)$\pm 5.95$(norm.) \\ 
0.314 & -- & 0.628 & 17.01 & 0.44(stat.) & $^{+1.21}_{-1.19}$(sys.)$\pm 4.82$(norm.) \\ 
0.628 & -- & 0.942 & 14.89 & 0.41(stat.) & $^{+1.00}_{-0.90}$(sys.)$\pm 4.22$(norm.) \\ 
0.942 & -- & 1.26 & 15.20 & 0.39(stat.) & $^{+0.80}_{-0.79}$(sys.)$\pm 4.30$(norm.) \\ 
1.26 & -- & 1.57 & 15.33 & 0.49(stat.) & $^{+0.70}_{-0.84}$(sys.)$\pm 4.34$(norm.) \\ 
\hline 
\multicolumn{6}{c}{$0.3 < \beta < 0.4$} \\ 
\hline 
0 & -- & 0.314 & 9.61 & 0.43(stat.) & $^{+0.76}_{-0.84}$(sys.)$\pm 2.72$(norm.) \\ 
0.314 & -- & 0.628 & 8.18 & 0.29(stat.) & $^{+0.59}_{-0.70}$(sys.)$\pm 2.32$(norm.) \\ 
0.628 & -- & 0.942 & 7.78 & 0.28(stat.) & $^{+0.53}_{-0.58}$(sys.)$\pm 2.20$(norm.) \\ 
0.942 & -- & 1.26 & 8.36 & 0.29(stat.) & $^{+0.62}_{-0.63}$(sys.)$\pm 2.37$(norm.) \\ 
1.26 & -- & 1.57 & 8.39 & 0.41(stat.) & $^{+0.58}_{-0.77}$(sys.)$\pm 2.38$(norm.) \\ 
\hline 
\multicolumn{6}{c}{$0.4 < \beta < 0.5$} \\ 
\hline 
0 & -- & 0.314 & 5.68 & 0.33(stat.) & $^{+0.63}_{-0.67}$(sys.)$\pm 1.61$(norm.) \\ 
0.314 & -- & 0.628 & 4.60 & 0.23(stat.) & $^{+0.57}_{-0.64}$(sys.)$\pm 1.31$(norm.) \\ 
0.628 & -- & 0.942 & 4.67 & 0.22(stat.) & $^{+0.47}_{-0.59}$(sys.)$\pm 1.33$(norm.) \\ 
0.942 & -- & 1.26 & 5.64 & 0.24(stat.) & $^{+0.49}_{-0.67}$(sys.)$\pm 1.60$(norm.) \\ 
1.26 & -- & 1.57 & 5.32 & 0.36(stat.) & $^{+0.45}_{-0.60}$(sys.)$\pm 1.51$(norm.) \\ 
\hline 
\multicolumn{6}{c}{$0.5 < \beta < 0.7$} \\ 
\hline 
0 & -- & 0.314 & 3.58 & 0.22(stat.) & $^{+0.44}_{-0.66}$(sys.)$\pm 1.02$(norm.) \\ 
0.314 & -- & 0.628 & 2.81 & 0.16(stat.) & $^{+0.42}_{-0.55}$(sys.)$\pm 0.80$(norm.) \\ 
0.628 & -- & 0.942 & 3.20 & 0.17(stat.) & $^{+0.44}_{-0.54}$(sys.)$\pm 0.91$(norm.) \\ 
0.942 & -- & 1.26 & 4.19 & 0.19(stat.) & $^{+0.44}_{-0.54}$(sys.)$\pm 1.19$(norm.) \\ 
1.26 & -- & 1.57 & 4.78 & 0.28(stat.) & $^{+0.43}_{-0.53}$(sys.)$\pm 1.36$(norm.) \\ 
\noalign{\hrule height 1.25pt} 
\end{tabular}
}
  \label{tab:phi}
\end{table}

\begin{table}
  \centering
  \caption{Results of the fit to the cross-section $\phiCross$ in bins
    of $\beta$. The fitted function is proportional to $(1+A \cosTwoPhi)$.
    The uncertainty includes both statistical and systematical contributions (see text).
    \vspace*{1cm}}
  \centering
{\renewcommand{\arraystretch}{1.35}
\begin{tabular}{r c r r@{$\pm$}p{1.5cm}}
\noalign{\hrule height 1.25pt} 
\multicolumn{3}{c}{ $\beta$ } & \multicolumn{2}{c}{$A$} \\ 
\hline
0.04 & -- & 0.15 & 0.256 & 0.030 \\
0.15 & -- & 0.3 & 0.130 & 0.028 \\
0.3 & -- & 0.4 & 0.053 & 0.045 \\
0.4 & -- & 0.5 & $-0.037$ & 0.054 \\
0.5 & -- & 0.7 & $-0.196$ & 0.070 \\
\noalign{\hrule height 1.25pt} 
\end{tabular}
}
  \label{tab:AvsBeta}
\end{table}

\clearpage

\newcommand{\figureScale}{1.1}
\newcommand{\figureTwoScale}{2}
\newcommand{\eFlowHeight}{0.9}
\newcommand{\diagHeight}{0.3}

\clearpage{}
\clearpage{}

\begin{figure}
\centering
\includegraphics[height=\eFlowHeight\textheight]{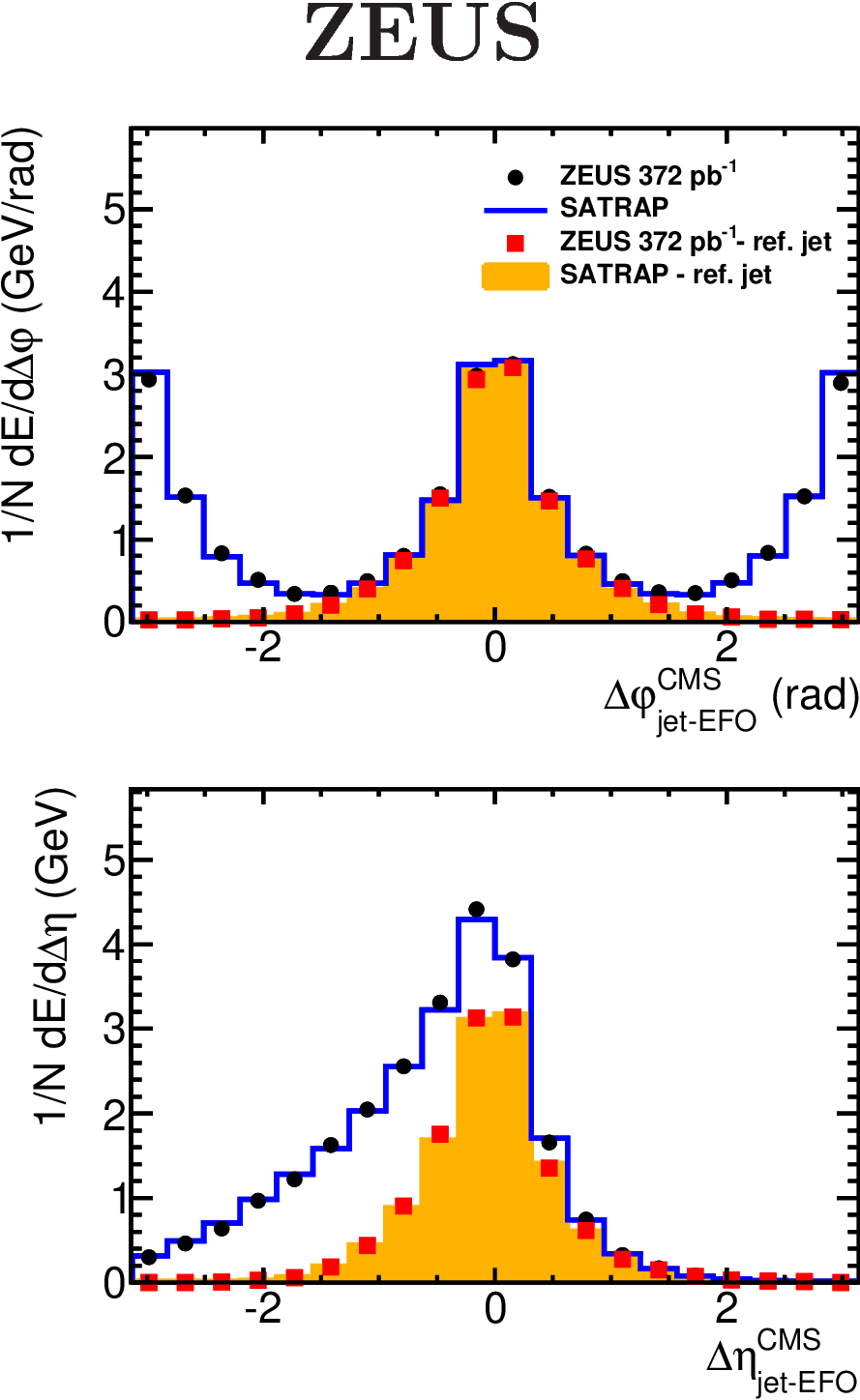}
\caption[]{The energy flow in the $\gamma^*$--${\rm I\!P}$ rest frame
  around the jet axis, averaged over all selected dijet events, is
  shown as a function of distances in azimuthal angle and
  pseudorapidity ($\Delta\varphi$ and $\Delta\eta$). In both cases, the
  energy flow is integrated over the full available
  range of the other variable. Data for both jets are shown as full
  dots. Statistical uncertainties are smaller than point markers. The energy flow of EFOs belonging to
  the reference jet only are shown as full squares, where the
  reference jet was chosen as the jet with positive $p_Z$
  momentum. Predictions of the SATRAP MC are shown as histograms. }
\label{fig_e_flow_cms}
\end{figure}

\begin{figure}
\centering
\includegraphics[height=\eFlowHeight\textheight]{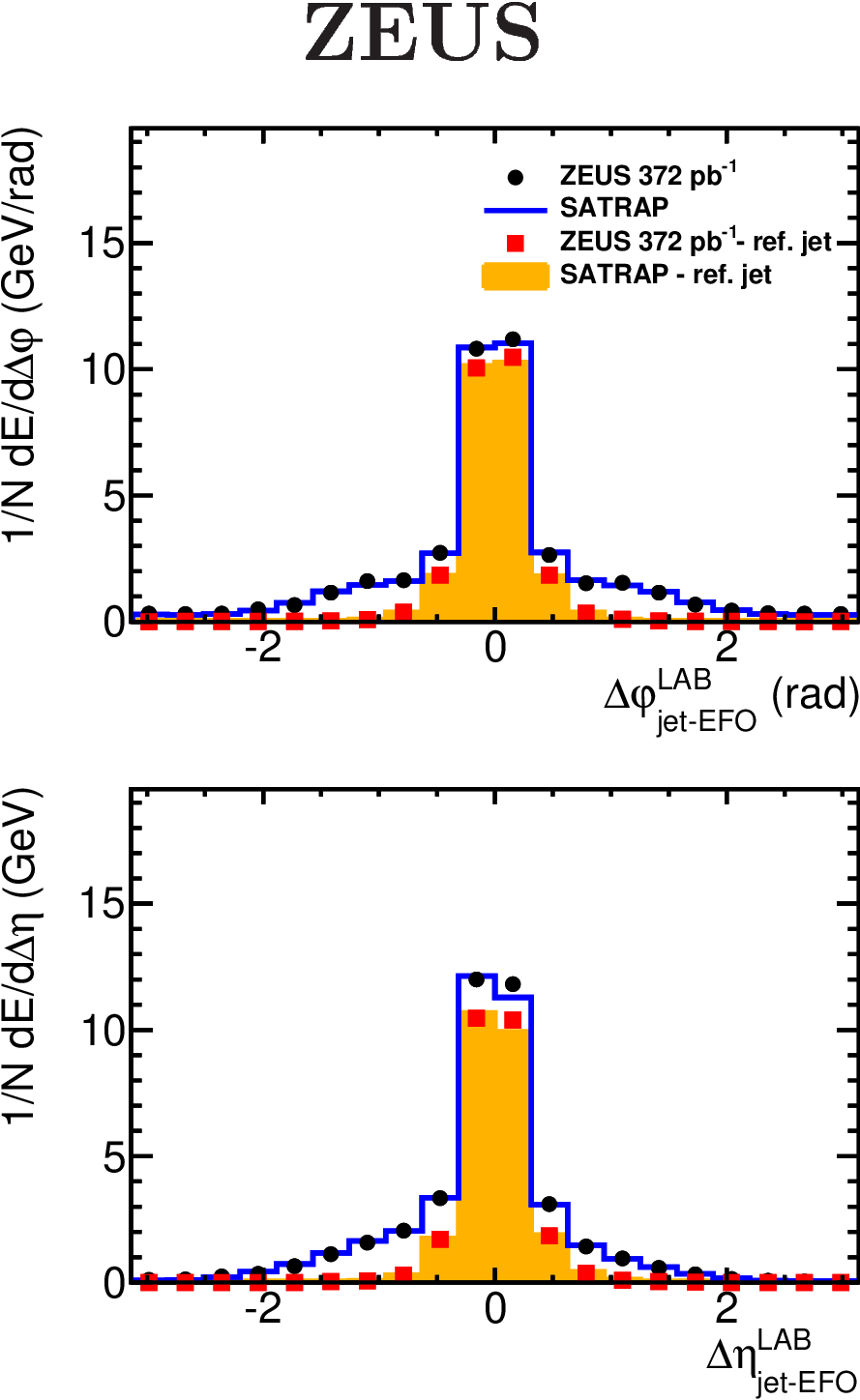}
\vspace{-0.15in}
\caption[]{The energy flow in the laboratory frame, around the jet
  axis, averaged over all selected dijet events, is shown as a
  function of distances in azimuthal angle and pseudorapidity
  ($\Delta\varphi$ and $\Delta\eta$). In both cases, the energy flow is
  integrated over the full available range of the other variable. Data
  for both jets are shown as full dots. Statistical uncertainties are
  smaller than point markers. The energy flow of EFOs belonging to the
  reference jet only are shown as full squares, where the reference
  jet was chosen as the jet with positive $p_Z$ momentum. Predictions
  of the SATRAP MC are shown as histograms.}
\label{fig_e_flow_lab}
\end{figure}

\clearpage{}

\clearpage{}

\clearpage{}

\end{document}